\documentclass[twoside,11pt]{article}

\usepackage{jmlr2e}
\usepackage{amsmath}
\usepackage{dsfont}
\usepackage{times}
\usepackage{multirow}
\usepackage{bm}
\usepackage{graphicx}
\usepackage{color}

\def\T{{ \mathrm{\scriptscriptstyle T} }}

\newcommand{\RN}[1]{%
  \textup{\uppercase\expandafter{\romannumeral#1}}%
}

\def\benr{\begin{eqnarray}}
\def\eenr{\end{eqnarray}}
\def\benrr{\begin{eqnarray*}}
\def\eenrr{\end{eqnarray*}}
\def\ben{\begin{equation}}
\def\een{\end{equation}}
\def\benn{\begin{equation*}}
\def\eenn{\end{equation*}}

\def\al{\alpha}
\newcommand{\cH}{{\cal H}}
\def\h{\hat}
\def\ep{\epsilon}
\def\iny{\infty}
\def\la{\lambda}
\def\ti{\tilde}
\def\G{\Gamma}
\def\si{\sigma}
\def\ka{\kappa}

\def\mbf{\mathbf}
\def\mbfa{\mathbf a}
\def\mbfb{\mathbf b}

\def\bH{\mathbf H}
\def\bX{\mathbf X}
\def\bY{\mathbf Y}
\def\bZ{\mathbf Z}

\def\bmbeta{\bm \beta}
\def\bmep{\bm \ep}
\def\bmPhi{\bm \Phi}
\def\bmtheta{\bm \theta}

\def\hDash{\bot\!\!\!\bot}
\def\nn{\nonumber}
\def\noi{\noindent}
\def\vs{\vskip}

\ShortHeadings{Local Average Estimate}{Peng, Xie, \& Zhao}
\firstpageno{1}

\begin{document}

\title{Fast Inference Procedures for Semivarying Coefficient Models via Local Averaging}

\author{\name Peng Heng \email hpeng@hkbu.edu.hk \\
       \addr Department of Mathematics\\
       Hong Kong Baptist University\\
       Kowloon Tong, Kowloon, Hong Kong
       \AND
       \name Xie Chuanlong$^*$ \email clxie0929@jnu.edu.cn \\
       \addr Department of Statistics\\
       Jinan University\\
       Huangpu Avenue 601, Guangzhou, China
       \AND
       Zhao Jingxin\email jessicazhao@wisers.com \\
       \addr Wisers AI Lab\\
       109-111 Gloucester Road, Wan Chai, Hong Kong
       }

\editor{.................}

\maketitle

\begin{abstract}
The semivarying coefficient models are widely used in the application of finance, economics, medical science and many other areas.
The functional coefficients are commonly estimated by local smoothing methods, e.g. local linear estimator.
This implies that one should implement the estimation procedure for hundreds of times to obtain an estimate of one function.
So the computation cost is very severe.
In this paper, we give an insight to the trade-off between statistical efficiency and computation simplicity, and proposes a fast inference procedure for semivarying coefficient model.
In our method, the coefficient functions are approximated by piecewise constants, which is a simple and rough approximation.
This makes our estimators easy to implement and avoid repeat estimation.  
In this work, we shall show that though these estimators are not asymptotically optimal, they are efficient enough for building further inference procedure. 
Furthermore, three tests are brought out to check whether certain coefficient is constant.
Our results clearly show that when the room for improving the asymptotic efficiency is limited, a proper trade-off between statistical efficiency and computation simplicity can be taken into consideration to improve the performance of the inference procedure.
\end{abstract}


\section{Introduction}

The semivarying coefficient model~\citep{zhang2002} is an extension of simple linear model, which assumes that some coefficients of a linear model are known to be functions of an index variable.
It has aroused interest of many researchers because of the dynamic coefficients.
With the varying coefficient part, this model is more flexible than a simple linear model and can express complicated relationship of the output against the inputs.
What's more, the parametric part makes it have good interpretability as a simple linear model.
For instance, in this age of big data, the E-business would collect many information from the consumers and make use of these information to do target promotion.
It will become more convincing if the association is allowed to change over time (or age).
Similarly, the semivarying coefficient model is successfully applied in economics, finance, epidemiology, medical science and many other areas.
The property of changing coefficient is quite appealing for analysis of nonlinear time series data, longitudinal data and survival data.

Let $Y$ be an output variable, and let 
\benn
\bX = (X_1, \ldots, X_p)^T, \quad \text{and} \quad \bZ = (Z_1, \ldots, Z_q)^T
\eenn
be input vectors with $p$-length and $q$-length respectively.
The semivarying coefficient model is in the form of
\ben\label{eq:semi-varying}
Y=\bX^T\mbfa(U) + \bZ^T \mbfb +\ep.
\een
where $U$ is the index variable, $\mbfb = (b_1, \ldots, b_q)^T$ , and $\mbf{a}(U) = (a_1(U), \ldots, a_p(U))^T$ is a smooth function.
A special case of the semivarying coefficient model is the varying coefficient model~\citep{hastie1993}, in which $\mbfb$ is a zero vector. It usually takes a form as
\ben\label{eq:varying}
Y = \bX^T \mbfa(U) + \ep.
\een
In the following, we review some related work about estimating $\mbfa(u)$ and $\mbfb$ via local or global smoothing methods.

We start with the varying coefficient model in (\ref{eq:varying}).
If $a_l(u)$ have the same degree of smoothness, \cite{hastie1993} proposed an estimation method with smoothing splines.
\cite{huang2002} and  \cite{huang2004} developed an another global smoothing method based on polynomial splines.
By choosing multiple smoothing parameters, their method works well when $a_l(u)$ have different degrees of smoothness.
On the other hand, since the varying coefficient model is locally approximated by a simple linear model, the kernel-based local smoothing estimators are also popular in the literature.
\cite{hoover1998} proposed a weighted local polynomial estimator and its asymptotic properties are derived by \cite{wu1998}.
This one-step estimator achieves a bias of $O(h^2)$ and a variance of $O((nh)^{-1})$ when all $a_l(u)$ possess the same degree of smoothness.
However, \cite{fan1999} pointed out that if this assumption does not hold, the optimal rate (\cite{fan1996}) can not be reached. So they proposed a two-step estimator.
Say different from the others, they require that the target coefficient function has bounded fourth derivatives.
With another tunable bandwidth $h_2$, the bias of the two-step estimator is of $O(h_2^4)$ and the variance is of $O((nh_2)^{-1})$.
So the two-step estimator can achieve the optimal rate of convergence $n^{-8/9}$. 

As to the semivarying coefficient model in (\ref{eq:semi-varying}), one can see that a good estimator of the constant coefficient vector $\mbfb$ will turn the problem into a varying coefficient model.
Then the remains can be solved by the methods we have mentioned above.
\cite{zhang2002} suggested to consider $\mbfb$ as functional too, e.g. $\mbfb(u)$, and then take average to get its final estimate.
The bias of their estimator of $\mbfb$ is of order $O(h^2)$ and the covariance matrix is of order $O(n^{-1})$.
We notice that this estimator is developed from a local estimator, which implies that the global property of $\bZ^T \mbfb$ in (\ref{eq:semi-varying}) is not fully utilized.
Then in \cite{fan2005}, a profile least-square estimator was put forward.
This estimator also has a bias $O(h^2)$ and a variance $O(n^{-1})$.
Besides, \cite{fan2005} have showed that unlike \cite{zhang2002}'s estimator, theirs is semiparametrically efficient.
But the cumbersome process of computing nuisance parameters is obvious a shortcoming.
To further reduce the estimation bias of $\mbfb$, \cite{xia2004} presented a semi-local least squares estimator.
The constant coefficient vector $\mbfb$ is estimated globally while the functional ones are estimated locally.
\cite{xia2004} have showed that their estimator has bias of $O(h^3)$ and the variance is $O(n^{-1})$.
Since the bias has been reduced, the undersmoothing is avoid.
However the computation burden is more heavier, since the size of the design matrix is increasing with $n^2$.
Alternatively, general series method can also be applied to semivarying coefficient model, see \cite{ahmad2005}.

Naturally we are also interested in the test problem that whether certain coefficient $a_l(u)$ is really varying.
The researchers have investigated many kinds of difference between the null and the alternative hypothesis to get the test statistics and the corresponding critical values.
\cite{fan2000} studied the deviation of the estimated coefficient function and the true coefficient function.
This test statistic is intuitional but involves many estimations for the unknown quantities.
Another approach is the log-likelihood ratio test, which should use bootstrap to the get the reject rules.
See \cite{cai2000}, \cite{cai2000b} and \cite{huang2002} for different estimators and data types.
\cite{fan2001} proposed the generalized likelihood ratio (GLR) tests and illustrated the idea with varying coefficient model in detail.
They have proved that the GLR tests are optimal and follow the Wilk's phenomena.

However, a growing concern of the computation cost has caused a vast number of studies to develop fast algorithms.
The estimators mentioned above need loads of computational work. 
What's worse, for model checking problem, one has to fit all $a_l(u)$, both under the null hypothesis and the alternative.
If the bootstrap is also used to determine the rejection region, the computation burden will be even heavier.
On the other hand, the room for improving the estimation efficiency is quite limited.
The optimal rate of the two-step estimator is already $n^{-8/9}$ and the asymptotic variance of $\mbfb$ is bounded by the semiparametric information matrix.
Thus, the excessive pursue for the estimation efficiency may gain little but make the method complicated and time consuming.
Therefore, a proper trade-off between the efficiency and the computational burden should be taken into consideration to improve the performance of the statistics methods.
Works about this topic seems scant and we make attempt to fill this void in this paper.

We come up with a local average method for estimating the varying coefficient model and the semivarying coefficient model.
The main idea of our method is to regard the varying coefficient function $a_l(u)$ as piecewise constant so that we can use least square to get a series of points estimators of $a_l(u)$. 
We call the proposed method as local average estimator.
In the following, we shall show that though the local average estimator is simple and rough, 
it provides a good base for further inference.
The local average estimator has three advantages. 
First, it sharply lighten the computation burden. 
The local linear or quadratic estimator only estimates the value of $a_l(u)$ at a given point $u_0$. So one should repeat the estimation procedure hundreds of times to obtain an estimate of the function $a_l(u)$. 
However, our method transforms the original model into a simple linear model, and directly estimate the values of $a_l(u)$ at a series of $u$.
Second, the bias of the local average estimator is very small, though it swells the variance.
Thus it provides necessary opportunity to develop adaptiveness to different degrees of smoothness. Third, this estimator can easily adapt to the semivarying coefficient model and result a global estimator of the constant coefficient $\mbfb$.

Our proposed methods introduce many parameters to model varying coefficient functions. 
Intuitively, the proposed methods will over-fit the varying coefficient functions in model (\ref{eq:varying}) and (\ref{eq:semi-varying}).
Thus our strategy is not widely used and does not follow the common suggestions about over-fitting or over-parameterization.
In this paper, we shall prove that though its variance is large, the bias of the local average estimator is small enough to build further inference procedures.
In Section~\ref{sec2}, we introduce the proposed estimators for varying coefficient model (\ref{eq:varying}) and semivarying coefficient model (\ref{eq:semi-varying}), and investigate their asymptotic properties.
Based on the local average estimator, we propose three tests in Section~\ref{sec3}, which can simplify the calculation and are flexible to apply on other models. A significant feature of the proposed tests is that they can only focus on certain coefficients and avoid complicated calculation caused by estimating nuisance coefficients.



\section{Assumptions}\label{Assumptions}

In this section, we present the needed assumptions in this paper. 

\noi (a1). $a_l'(\cdot)$ and $a_l''(\cdot)$ are continuous and bounded for $l=1,...,p$.
\vs .1cm
\noi (a2). The function $a_p$ has continuous and bounded fourth derivative.
\vs .1cm
\noi (X). $\|\mbf{X}\|^2<\iny$, $\|\mbf{Z}\|^2<\iny$, and $\G(u, I)=\text{E}[(\sum_{i=1}^I X_i X_i^T)^{-1}|U=u]$ exists and is continuously differentiable with respect to any $u$ in the support of $U$.
\vs .1cm
\noi ($\ep$1).  $\text{E}[\ep| U, X] =0$, $\text{Var}[\ep|U,X] = \si^2$.
\vs .1cm
\noi ($\ep$2). $\text{E}[\ep^4]=\mu_4<\iny$.
\vs .1cm
\noi (U). The density function $f_U$ of $U$ has bounded first-order derivative and satisfies
\[
0<\delta \leq \inf_{u}f_U(u)\leq\sup_{u}f_U(u)<\iny.
\]
\vs .1cm
\noi (K). The function $K(u)$ is a symmetric density function with a compact support.
\vs .1cm
\noi (I). The group size $I$ is a small integer such that  $I/n\to 0$.
\vs .1cm
\noi (h1). Denote $h = h_n$ is a sequence of bandwidths, and assume $h\rightarrow0$,  $nh\rightarrow\iny$ as $n \rightarrow \iny$.
\vs .1cm
\noi (h2). $h\to 0$, $nh^{3/2}\to \iny$ as $n\to \iny$.

\section{Local Average Estimator}\label{sec2}

In this section, we introduce the local average estimator and illustrate how to build the estimation procedure of the varying coefficient model in (\ref{eq:varying}) and the semivarying coefficient model in (\ref{eq:semi-varying}) via the local average estimator. 
Further more, we systematically investigate the large sample properties of the proposed methods.  

\subsection{Varying coefficient model}\label{sec21}

We first consider the varying coefficient model. Assume that the collected data is $\{(U_i, \bX_i, Y_i), i=1,\cdots,n\}$. 
In the beginning, we sort the samples according to $U_i$ in an ascending order. Denote $U_{(1)}\leq U_{(2)}\leq ... \leq U_{(n)}$. Then divide them into $k$ groups with $I$ samples in each group, where $I$ is a fixed integer and $n=Ik$. (In practise, the possible remainders are removed out. Since $I$ is small enough, the number of the removed samples is negligible.)

Denote $\bX$ and $Y$ corresponding to $U_{(i)}$ as $\bX_{(i)}$ and $Y_{(i)}$. 
Thus the $j$-th observation in $i$-th group is $(U_{(iI -I +j)},\bX_{(iI -I +j)},Y_{(iI -I +j)})$ and
\benn
Y_{(iI -I +j)} = \bX_{(iI -I +j)}^T \mbfa(U_{(iI -I +j)}) +\ep_{(iI -I +j)}, \quad i=1,\ldots,k, \quad j=1,\ldots,I,
\eenn
where $\bX_{(iI -I +j)} = (X_{(iI -I +j),1}, \ldots, X_{(iI -I +j),p})^T$ and $\ep_{(iI -I +j)}$ is the corresponding error for the $j$-th observation in $i$-th group.
We assume, for each $1\leq i \leq k$,
\ben\label{lgca}
\mbfa(U_{(iI -I +1)}) = \cdots = \mbfa(U_{(iI)}) \equiv \mbfa_i = \mbfa(\bar{U}_{i\cdot}),
\een
where $\bar{U}_{i\cdot} = \sum_{j=1}^I U_{(iI -I +j)}/I$.
Let $\ep^*_{(iI -I +j)} = Y_{(iI -I +j)} - \bX_{(iI -I +j)}^T \mbfa_i$.
To proceed further, we denote
\benr\label{def1}
&& \mbfa_i = (a_1(\bar{U}_{i\cdot}), a_2(\bar{U}_{i\cdot}), \ldots,a_p(\bar{U}_{i\cdot}))^T, \quad \mbfa = (\mbfa_1^T, \mbfa_2^T, \ldots, \mbfa_k^T)^T, \\
&& \bmep^*_i = (\ep^*_{(iI -I +1)}, \ep^*_{(iI -I +2)}, \ldots, \ep^*_{(iI)})^T, \quad \bmep^* = (\bmep_1^{*T}, \bmep^{*T}_2, \ldots, \bmep^{*T}_k)^T, \nn \\
&& \bY^*_i=(Y_{(iI -I +1)}, Y_{(iI -I +2)}, \ldots, Y_{(iI)})^T,\quad \mathds{Y} = (\bY_1^{*T}, \bY_2^{*T}, \ldots, \bY_k^{*T})^T, \nn \\
&& \bX^*_i=(\bX_{(iI -I +1)}, \bX_{(iI -I +2)}, \ldots, \bX_{(iI)})^T, \quad \mathds{X} = diag(\bX^*_1, \bX^*_2, \ldots, \bX^*_k). \nn
\eenr
Thus, we know
\ben\label{varying1}
\bY_i = \bX^*_i\mbf{a}_i+ \bmep^*_i, \,\,\,\, \text{and} \,\,\,\, \mathds{Y} = \mathds{X}\mbfa+\bmep^*.
\een
Now we get the local average estimator
\benr\label{primary}
\h{\mbf{a}}&=&(\h a_1(\bar U_{1\cdot}), \cdots, \h a_p(\bar U_{1\cdot}), \h a_1(\bar U_{2\cdot}), \cdots, \h a_p(\bar U_{2\cdot}), \cdots, \h a_1(\bar U_{k\cdot}), \cdots, \h a_p(\bar U_{k\cdot}))^T \nn \\
 &=&(\mathds{X}^T \mathds{X})^{-1} \mathds{X}^T \mathds{Y}
\eenr
For each $a_l(\cdot)$, from $\h{\mbf{a}}$, relevant estimators $\h a_l(\bar U_{1\cdot}),\h a_l(\bar U_{2\cdot}), ..., \h a_l(\bar U_{k\cdot})$ are acquired.
The following lemma states the large sample properties of these point estimators $\h a_l(\bar{U}_{i\cdot}= u)$ obtained by local averaging.
Its proof is postponed in to the {\bf Appendix}.
\begin{lemma}\label{th:localaverage}
Suppose (a1), (X), (U), (I) hold. Then for $\h a_l(\bar{U}_{i\cdot}= u)$ in (\ref{primary}), we have
\benrr
&&\text{E}[\h a_l(\bar{U}_{i\cdot}= u)]=a_l(u)+ O(\frac{\log n}{n}), \\
&&\text{Var}[\h a_l(\bar{U}_{i\cdot} = u )]=e_{l,p}^T \G(u, I) e_{l,p}\si^2, \quad l=1,\ldots p, i=1,\ldots k.
\eenrr
\end{lemma}
This lemma implies that the local average estimator can be rewritten as random sample from a nonparametric model
\begin{equation}\label{simple}
\h a_l(\bar{U}_{i\cdot})=a_l(\bar{U}_{i\cdot})+ \eta_i + O_p(\frac{\log n}{n}), \,\,\, i=1,\ldots,k
\end{equation}
where $\text{E}[\eta_i|\bar{U}_{i\cdot}] = 0$ and $\text{Var}[\eta_i|\bar{U}_{i\cdot}] = e_{l,p}^T\G(\bar{U}_{i\cdot}, I)e_{l,p}$.

\begin{remark}
The piecewise constant approximation in (\ref{lgca}) transforms one coefficient function $a_l(u)$ into a $k$-length vector. 
Note that $k=n/I$ is at the same order of $n$. 
This implies that comparing to the sample size $n$, the number of parameters is large.
According to {\bf Lemma}~(\ref{th:localaverage}), one can see that the local average estimator $\h a_l(u)$ is inconsistent, because its variance does not converge to zero as $n$ tends to infinity.
On the other hand, the bias is at the order of $\ln(n)/n$, which is much smaller than the bias of a local smoothing estimate.
These confirm that over-fitting exists.

\end{remark}

Now we can use local smoothing methods to further estimate $a_l(u)$, and adaptively choose bandwidth $h$ and other parameters according to the smoothness of $a_l(u)$. In this paper, we take $l = p$ for example and adapt the local polynomial smoothing.
For given $u$, denote
\[
\mbf{\bar U}=\begin{pmatrix}
            1&(\bar U_{1\cdot}-u)&\cdots&(\bar U_{1\cdot}-u)^3\\
            \vdots&\vdots& &\vdots\\
            1&(\bar U_{k\cdot}-u)&\cdots&(\bar U_{k\cdot}-u)^3
            \end{pmatrix},
\]
and put
\benrr
\h{\mbfa}_p &=& (\h a_p(\bar U_{1\cdot}), \h a_p(\bar U_{2\cdot}), \cdots, \h a_p(\bar U_{k\cdot}))^T, \\
\mbf{\bar W} &=& diag(K_h(\bar U_{1\cdot}-u),\ldots, K_h(\bar U_{k\cdot}-u)),
\eenrr
where $K$ is a kernel function and $K_h = K(./h)/h$.
Then the further estimator of $a_p(u)$ can be obtained by
\ben\label{Step2Est}
\ti a_p(u) = e^T_{1,4}(\mbf{\bar U}^T \mbf{\bar W} \mbf{\bar U})^{-1} \mbf{\bar U}^T \mbf{\bar W} \h{\mbf{a}}_p.
\een
Next we shall show that $\ti a_p(u)$ can converge to $a_p(u)$ at the optimal rate $n^{-8/9}$.
To proceed further, we denote
\begin{align*}
\xi_i=\int t^iK(t)dt, \quad \text{and}\quad \nu_i=\int t^iK^2(t)dt.
\end{align*}
Now we are ready to state
\begin{theorem}\label{thm:vary1}
Suppose (a1), (a2), (X), ($\ep$1), (U), (K), (I) and (h1) hold. 
Then for given $u$,  the asymptotic bias of $\ti a_p(u)$ in (\ref{Step2Est}) is
\[
\text{bias}[\ti a_p(u)] = \frac{1}{4!}\frac{\xi_4^2-\xi_2\xi_6}{\xi_4-\xi_2^2}a_p^{(4)}(u)h^4+o_p(h^4)
\]
and the asymptotic variance of $\h a_p(u)$ is given by
\[
\text{var}[\ti a_p(u)] = \frac{(\xi_4^2\nu_0-2\xi_4\xi_2\nu_2+\xi_2^2\nu_4)\si^2I}{nhf_U(u)(\xi_4-\xi_2^2)^2} e_{p,p}^T\G(u, I) e_{p,p}+o_p(\frac{1}{nh}).
\]
where $a_p^{(4)}(u)$ is the $4$-th order derivative of $a_p(u)$ with respect to $u$.
\end{theorem}
The asymptotic bias and variance of our estimator have the same order $O(h^4)$ and $O((nh)^{-1})$ to those of  \cite{fan1999}'s two step estimator.
Thus the MSE of our estimator can achieves the optimal rage of convergence $n^{-8/9}$ when $h$ is taken of order $n^{-1/9}$.
Comparing to \cite{fan1999}'s two step estimator, the proposed estimator $\ti a_p(u)$ has the same order of asymptotic variance $O((nh)^{-1})$. 
For the asymptotic bias, our estimator is of $O(h^4)$ as well, but the formula is more concise since we do not have the term dominated by the initial bandwidth. 
Also, the conditional MSE of the local average estimator can achieves the optimal rage of convergence $n^{-8/9}$ when $h$ is taken of order $n^{-1/9}$. 
Other theoretical advantages of \cite{fan1999}'s two step estimator also hold in the local average estimator. 
For example, the estimators has the same optimal convergent rate as in the ideal situation where $a_1,\ldots,a_{p-1}$ are known.

The following theorem  provides the asymptotic properties of the estimator $\ti a_p(u)$ in the case that the objective coefficient $a_p(u)$ shares the same smoothness with others.
That is to say, $a_p(u)$ has continuous and bounded second derivative.
So in the local polynomial smoothing step (\ref{Step2Est}), we applied a linear fit.

\begin{theorem}\label{thm:vary2}
Suppose (a1), (X), ($\ep$1), (U), (K), (I) and (h1) hold. 
Then for given $u$, the asymptotic bias of $\ti a_p(u)$ in (\ref{Step2Est}) is
\[
\text{bias}[\ti a_p(u)]=\frac{1}{2}\xi_2 a_p''(u)h^2+o_p(h^2)
\]
and the asymptotic variance of $\h a_p(u)$ is given by
\[
\text{var}[\ti a_p(u)]=\frac{\nu_0\si^2 I}{nhf_U(u)}e_{p,p}^T \G(u,I) e_{p,p}+o_p(\frac{1}{nh}).
\]
\end{theorem}

Now the asymptotic bias is of $O(h^2)$ and the asymptotic variance is of $O((nh)^{-1})$.
What's more, the asymptotic result is the same as that of the one-step estimator \citep{hoover1998}, and the bias is one term less compared with the two-step estimator \citep{fan1999}.
In other words, the local average estimator performs as well as the one-step estimator when there is no smoothness difference among the coefficient functions $a_l(u)$, $l=1, \ldots, p$.
Notice that we apply local polynomial smoothing in the second step and the above asymptotic properties are all based on this setting. 
Obviously, the asymptotic results will change if different smoothing method is chosen. 
However, the local average estimators in (\ref{primary}) are asymptotically biased and their variances have explicit forms.
What's more, those estimators are independent.
Therefore, common-used nonparametric regression techniques are available for the smoothing step and their asymptotic properties will not be skewed.
In this way, our proposed estimator is very flexible .
Prior information about the objective functional coefficients could be fully utilized with various smoothing methods.

\subsection{Semivarying coefficient model}\label{sec22}

The local average estimator can be readily extended to the semivarying coefficient model in (\ref{eq:semi-varying}). 
Denote the samples as $\{ (U_i, \bX_i, \bZ_i, Y_i),\,\,i=1,\ldots,n \}$. 
After ordering and grouping these samples according to $U$, we index the $j$-th observation in $i$-th group as
\benn
(U_{(iI -I +j)}, \bX_{(iI -I +j)}, \bZ_{(iI -I +j)}, Y_{(iI -I +j)}), \,\,\, i =1,\ldots, k\,\,\,\text{and}\,\,\, j =1,\ldots,I.
\eenn
To proceed further, we denote $\bmPhi=(\mathds{X},\mathds{Z})$ and $\bmtheta = (\mbfa^T, \mbfb^T)^T$ where
\ben\label{def2}
\mathds{Z}=(\bZ_1^{*T}, \bZ_2^{*T}, ..., \bZ_k^{*T})^T, \quad  \bZ^*_i = (Z_{(iI -I +1)}, Z_{(iI -I +2)}, ..., Z_{(iI)})^T
\een
and $\mathds{X}$, $\mbfa$ are similar to those in (\ref{def1}).
Then we can write the model as
\ben\label{varying2}
\mathds{Y} = \bmPhi \bmtheta +\bmep^*,
\een
where $\ep^*_{(iI -I +j)}= Y_{(iI -I +j)} - \bX_{(iI -I +j)}^T \mbfa_i - \bZ_{(iI -I +j)}^T \mbfb$ and
\benn
\bmep^*=(\bmep_1^{*T},\bmep^{*T}_2,\ldots,\bmep^{*T}_k)^T, \quad \bmep^*_i=(\ep^*_{(iI -I +1)}, \ep^*_{(iI -I +2)},\ldots,\ep^*_{(iI)})^T.
\eenn
Therefore the local average estimator of the parameter $\mbf{b}$ is given by
\ben\label{Step2EstSemi}
\h\mbfb=(\mbf{0}_{1\times kp}, \mbf{1}_{1\times q})(\bmPhi^T\bmPhi)^{-1}\bmPhi^T \mathds{Y}.
\een
In the following, one can see that $\h\mbfb$ is still a $\sqrt n$-consistent estimator of $\mbfb$.
For the varying coefficient part, either a back substitution or continuation with classical smoothing rebuild is available.

\begin{theorem}\label{thm:semi}
Suppose the assumptions (a1), (X), ($\ep$1), (U), (K), (I) hold. Then
\[
\sqrt{n}(\h\mbfb - \mbfb)\Rightarrow \text{N}(\mbf{0},\si^2\Sigma^{-1})
\]
where $\mbfb$ is the local average estimator in (\ref{Step2EstSemi}) and
\[
\Sigma= \text{E}(\bZ\bZ^T)-\text{E}\bigg\{\text{E}\Big [(\frac{1}{I}\sum_{j=1}^I \bZ_j\bX_j^T)(\frac{1}{I}\sum_{j=1}^I \bX_j\bX_j^T)^{-1}(\frac{1}{I}\sum_{j=1}^I \bX_j\bZ_j^T)|U_1,...,U_I \Big]\bigg\}.
\]
\end{theorem}

Theorem \ref{thm:semi} states the asymptotic normality of the local average estimator for the constant coefficient. One can find that the group size $I$ effects the asymptotic variance. If we consider the case when $p=1$ and $X=1$, then the model in (\ref{eq:semi-varying}) will turn into
\[
Y = a(U) + Z^T\mbf{b} + \ep.
\]
By Theorem \ref{thm:semi}, the asymptotic variance will become $\frac{I}{I-1}\si^2\widetilde{\Sigma}^{-1}$, with
\[
 \widetilde{\Sigma}= \text{E}[\{Z - \text{E}(Z|U)\}\{Z - \text{E}(Z|U)\}^T].
\]
This is consistent with the result of \cite{cui2017}. However, notice that \cite{bickel1993} have shown that $\si^2\widetilde{\Sigma}^{-1}$ is the semiparametric information bound. 
This implies that our local estimator doesn't reach the semiparametric efficient bound for general varying-coefficient partially linear model. 
This inefficiency is the expense for the computation simplicity.

However, if only $\mbfb$ is of interested, to estimate $\mbfa(u)$ will cause needless computation cost.
Thus it is a waste of computing power.
To deal with this problem, the local average estimator can be rewritten as a projection-based approach, which directly estimate $\mbfb$ without computing $\h \mbfa(u).$ 
The original problem is to find a vector $\mbfb$ and a function $\mbfa(u)$ to minimize the error function
\benr\label{E0}
\text{E}_0(\mbfa(.), \mbfb) = \sum_{i=1}^n \left(Y_i - \bX_i^\T \mbfa(U_i) - \bZ_i \mbfb\right)^2.
\eenr
Then, by grouped local constant approximation (\ref{lgca}), we have
\benrr
\text{E}_0(\mbfa(.), \mbfb) \approx \text{E}(\mbfa, \mbfb) = \sum_{i=1}^k \|\bY^*_i - \bX^*_i \mbfa_i - \bZ^*_i \mbfb\|^2
\eenrr
where $\|.\|$ is Euclidean norm, $\mbfa$, $\bY^*_i$ and $\bX^*_i$ are defined in (\ref{def1}), and $\bZ^*_i$ is defined in (\ref{def2}).
The estimates of $\mbfb$ and $\mbfa$ are given by solving
\benr\label{esteqb}
\sum_{i=1}^k \bZ^{*T}_i (\bY^*_i - \bX^*_i \mbfa_i - \bZ^*_i \mbfb) = 0
\eenr
and, for any $i=1,\ldots,k,$
\benr\label{esteqa}
\bX^{*T}_i(\bY^*_i - \bX^*_i \mbfa_i - \bZ^*_i \mbfb) =0.
\eenr
For any given $\mbfb$, (\ref{esteqa}) implies
\benrr
\h{\mbfa}_i = (\bX_i^{*T} \bX_i^*)^{-1} \bX_i^{*T}(\bY^*_i - \bZ^*_i \mbfb), \quad i=1,\ldots,k.
\eenrr
Plug these equations in to (\ref{esteqb}), we have
\benrr
\sum_{k=1}^K \bZ_i^{*T} (\mbf{I}_I - \bH_i) (\bY^*_i - \bZ^*_i \mbfb) =0,
\eenrr
where $\bH_i = \bX^*_i(\bX_i^{*T} \bX^*_i)^{-1}\bX_i^{*T}$ is the projection matrix for the column space of $\bX_i^*$, and $\mbf{I}_I$ is a $I\times I$ identity matrix.
Denote
\benrr
 \mathds{P} = \mbf{I}_n - \bH_1 \oplus \bH_2 \oplus \cdots \oplus \bH_k,
\eenrr
where $\oplus$ represents direct sum and $\mbf{I}_n$ is a $n \times n$ identity matrix here.
Then the estimation equation (\ref{esteqb}) and the error function $\text{E}(\mbfa, \mbfb)$ can be rewritten as
\benr\label{PLS}
\mathds{Z}^\T \mathds{P} (\mathds{Y} - \mathds{Z}\mbfb) = 0,\quad \text{and} \quad \text{E}(\mbfb) = \|\mathds{P} (\mathds{Y} - \mathds{Z} \mbfb)\|^2,
\eenr
and the estimate of $\mbfb$ is
\benr\label{estb}
\h{\mbfb} = (\bZ^\T \mathds{P} \bZ)^{-1}\bZ^\T \mathds{P} \bY.
\eenr
It is easy to see that $\hat{\mbfb}$ is the least square estimate with $\mathds{P}\mathds{Y}$ by $\mathds{P}\mathds{Z}$, and satisfies that
$
\mathds{P} (\mathds{Y} - \mathds{Z}\mathds{b}) \hDash \mathds{P}\mathds{Z}.
$
So we call $\hat{\mbfb}$ as local average projection estimator (LAPE) in the following.
\cite{zhao2015} proposed an iterated two-stage projection-based estimation for semivarying coefficient model.
The projection step removes $\bX_i^\T \mbfa(U_i)$ in (\ref{E0}) without ranking data, and their projection matrix is
$
\ti{\mathds{P}} = \mbf{I}_n - \bX(\bX^\T \bX)^{-1}\bX^\T
$
where
\benrr
\bX = \left(
        \begin{array}{cccc}
          \bX_1^\T &        &        &        \\
                 & \bX_2^\T &        &        \\
                 &        & \ddots &        \\
                 &        &        & \bX_n^\T \\
        \end{array}
      \right).
\eenrr
In other words, it projects $\bX_i^\T \mbfa(U_i)$ in (\ref{E0}) to the orthogonal space of $span(\bX_i)$, which is always 0 for any $\bX_i$.
Actually, this projection step is a special local averaging procedure with $I=1$.
Here we assume $I \geq d$ to make sure the identibility of $\hat{\mbfa}$.

\section{Hypothesis Testing}\label{sec3}

\subsection{Test Statistics}\label{sec31}
In this section, we use the local average estimator (\ref{primary}) to build three test statistics to deal with the model checking problem.
The testing problem of interest here is:
\ben\label{eq:hypothesis}
\cH_0:  a_p(u)=c,\quad \text{vs}\quad \cH_1: a_p(u)\neq c
\een
where $c$ is an unknown constant.
More specifically, the hypothesis test problem should be
\benrr
&& \cH_0: P(a_p(U)=c) = 1, \quad \text{for some constant $c$;} \\
&& \cH_1: P(a_p(U)\neq c) < 1, \quad \text{for all constant $c$.}
\eenrr
For simplicity, we will write this hypothesis test in the form of (\ref{eq:hypothesis}).
Recall that in the local average estimation process, we have transformed the varying coefficient part into a simple nonparametric model (\ref{simple}):
\benn
\h a_p(\bar{U}_{i\cdot})=a_p(\bar{U}_{i\cdot})+\eta_i + O_p(\frac{\log n}{n}), \quad i=1,\ldots,k
\eenn
with $E[\eta_i|\bar{U}_{i\cdot}]=0$ and $Var[\eta_i|\bar{U}_{i\cdot}]=e_{p,p}^T\G(\bar{U}_{i\cdot}, I)e_{p,p}$.
Then
some classical tests are available to check (\ref{eq:hypothesis}).
Note that the nonparametric model (\ref{simple}) is heteroscedastic, so we have to be careful when choosing the tests.

Firstly, we propose a moment-based test according to \cite{zheng1996}'s test.
Let $e_i = a_p(\bar{U}_{i\cdot}) -c$.
Then  $E\{E[e_i|\bar{U}_{i\cdot}] e_i\}$ should be closed to zero under $\cH_0$ and converge to a positive scalar when $\cH_1$ is true.
Hence our first test statistics is defined by
\[
T_1=\frac{1}{k(k-1)}\sum_{i=1}^{k}\sum_{j\neq i}^{k}\frac{1}{h}K(\frac{\bar{U}_{i\cdot}-\bar{U}_{j\cdot}}{h})\h e_i\h e_j,
\]
where $\h e_i=\h a_p(\bar{U}_{i\cdot})-\h c$ and $\h c=\sum_{i=1}^k \h a_p(\bar{U}_{i\cdot})/k$.
If the conditional variance $Var[\eta_i|\bar{U}_{i\cdot}=u]$ is known or can be estimated efficiently, we can also apply the generalized likelihood ratio(GLR, \cite{fan2001}) test to this problem. 
Then test statistics is
\[
T_2=\frac{n}{2I}\log\frac{\sum_{i=1}^k(\h a_p(\bar{U}_{i\cdot})-\h c)^2} {\sum_{i=1}^k(\h a_p(\bar{U}_{i\cdot})-\ti m_h(\bar{U}_{i\cdot}))^2}
\]
where $\ti m_h(\bar{U}_{i\cdot})$ is a nonparametric estimator of $a_p(\bar{U}_{i\cdot})$, for example, the local linear estimator or the Nadaraya-Watson estimator.

\begin{remark}
It is easy to see that $\h a_p(u)$ is a biased estimator of $a_p(u)$ and the bias term is of order $O(\log n/n)$.
Thus $\h e_i$ and $\h c$ are biased.
In the proof, we shall show that compare to the consistency rate of $T_1$ and $T_2$, the bias terms are asymptotically negligible.
In Section~\ref{sec32}, we shall show that the asymptotic properties of $T_1$ and $T_2$ are quite similar to those of the classical tests.
\end{remark}

\begin{remark}
In the $T_1$ test, we only consider one functional coefficient $a_p(\cdot)$ and construct the test statistic by $\h a_p(.)$, which is an rough and well-obtained estimator of $a_p(\cdot)$.
Similarly in $T_2$, we only need to estimate $a_p(\cdot)$, i.e. $\h c$ and $\ti m_h(\cdot)$.
If we directly apply the GLR test to the original varying coefficient model, we have to estimate other $2(p-1)$ uninterested functional coefficients. 
Hence in these tests, the computation cost has been sharply lessened after using the local average estimator.
\end{remark}

Notice that the GLR test is only based on the residual square, we may ignore all the function estimation if we can directly and efficiently estimate the residual variance .
Remind that in the local average estimator (\ref{primary}), we get the point estimators for the functional coefficients. Thus we can substitute these point estimators back to the varying coefficient or semivarying coefficient model to estimate the residual errors.

When $\cH_1$ is true, the alternative model is a varying coefficient model as (\ref{varying1}).
Therefore, by the local average estimator, the sum of residual square can be written as
\[
\widehat{\text{RSS}}_1=\mathds{Y}^T \mathds{P}_1\mathds{Y}
\]
where $\mathds{P}_1=\mbf{I}_n-\mathds{X}(\mathds{X}^T\mathds{X})^{-1}\mathds{X}^T$, and $\mathds{X}$, $\mbf{Y}$ are defined in (\ref{def1}).
The estimation of $\text{RSS}_0$ under null hypothesis is similar.
If $\cH_0$ is true, the last coefficient $a_p(u)$ is constant.
This implies that the null model is a semivarying coefficient model (\ref{varying2})
where $\bZ = X_p$ with $q=1$ and  $\bX=(X_1,\ldots, X_{p-1})^T$.
Therefore the sum of residual square under the null hypothesis can be estimated as
\[
\widehat{\text{RSS}}_0=\mathds{Y}^T\mathds{P}_0 \mathds{Y}
\]
where $\mathds{P}_0=\mathds{I}_n-\bmPhi (\bmPhi^T \bmPhi)^{-1} \bmPhi^T$, and $\bmPhi=(\mathds{X},\mathds{Z})$ with $\mathds{X}$ and $\mathds{Z}$ in (\ref{def2}).
Then the third test statistic is
\[
T_3=\frac{n}{2}\frac{\widehat{\text{RSS}}_0-\widehat{\text{RSS}}_1}{\widehat{\text{RSS}}_1}=\frac{n}{2}\frac{\mathds{Y}^T\mathds{P}_0\mathds{Y} - \mathds{Y}^T\mathds{P}_1 \mathds{Y}}{\mathds{Y}^T\mathds{P}_1\mathds{Y}}.
\]

\begin{remark}
For the $T_3$ test, the wanted functional coefficient has been carefully estimated under the null hypothesis and the nuisance coefficients are just simply approximated by local average estimator. This merit makes the proposed test very attractive when the testing problem is focusing on individual coefficient.
\end{remark}

\begin{remark}
The test statistic $T_3$ is a linear approximation of $(n/2) \log\{\widehat{\text{RSS}}_0/\widehat{\text{RSS}}_1\}$, which is the test statistic of the GLR test. 
However the difference between $(n/2)\log\{\widehat{\text{RSS}}_0/\widehat{\text{RSS}}_1\}$ and $T_3$ is not asymptotically negligible after timing $h^{-1/2}$, which is the standard sequence of $T_3$. So we must take the nonlinear part of the Taylor expansion of $\log\{\widehat{\text{RSS}}_0/\widehat{\text{RSS}}_1\}$ into consideration.
\end{remark}

\subsection{Limit null distributions}\label{sec32}

In this section, we will establish the limit null distributions of the proposed tests in Section 3. To state the following theorems, we need more notations as
\benrr
&& \dot{\bX}_{(iI -I +j)} = (X_{(iI -I +j),1}, X_{(iI -I +j),2}, \ldots, X_{(iI -I+j),p-1} ) ^T, \\
&&\Psi_n=\sum_{i=1}^k (B_i-C_i^T A_i^{-1}C_i)^{-2}\sum_{j=1}^I (C_i^TA_i^{-1}\dot{\bX}_{(iI -I +j)} -X_{(iI -I +j), p})^4, \\
&&\ka_1=K(0)-\frac{1}{2}\int K^2(t) dt, \quad \ka_2=\int\{K(t)-\frac{1}{2}K\ast K(t)\}^2 dt,
\eenrr
where $K\ast K$ denotes the convolution of $K$ and
\benrr
A_i=\sum_{j=1}^I\dot{\bX}_{(iI -I +j)}\dot{\bX}_{(iI -I +j)}^T, \quad B_i=\sum_{j=1}^I X_{(iI -I +j), p}^2, \quad C_i=\sum_{j=1}^I \dot{\bX}_{(iI -I +j)} X_{(iI -I +j), p}.
\eenrr
Combining the Lemma \ref{th:localaverage} and \cite{zheng1996}, we can get the limit null distribution of $T_1$.
\begin{theorem}\label{thm:T1}
Suppose the assumptions (a1), (X), ($\ep$1),($\ep$2), (U), (K) and (h1) hold. 
If the null hypothesis $\cH_0$ is true, $n h^{1/2}T_1\Rightarrow N(0,\si^2_1)$ where
\begin{equation*}
\si^2_1=2\si^4 I^2 \int K^2(s)ds\cdot\int (e_{p,p}^T \G(u, I)e_{p,p})^2f_U(u)d(u).
\end{equation*}
\end{theorem}
Then the standardized test statistic is given by $V_1 \equiv nh^{1/2}T_1/\h \si_1,$
where $\widehat{\si^2_1}$ is a consistent estimator of $\si^2_1$:
\begin{equation*}
\widehat{\si^2_1}=\frac{2I^2}{k(k-1)}\sum_{i=1}^k\sum_{j\neq i}^k\frac{1}{h}K^2(\frac{\bar U_{i\cdot}-\bar U_{j\cdot}}{h})\h e_i^2\h e_j^2.
\end{equation*}
By the Slutsky's theorem, $V_1 \Rightarrow N(0,1)$. Hence the $T_1$ test rejects $\cH_0$ whenever
$V_1 > z_\al $, where $z_{\al}$ is the upper $100(1-\al)\%$ quantile of the standard normal distribution.

\begin{theorem}\label{th:GLR1}
Suppose the assumptions (a1), (X), ($\ep$1), ($\ep$2), (U), (K) and (h2) hold. 
If the null hypothesis $\cH_0$ is true, we have $r_n T_2 \Rightarrow \chi^2_{a_n}$ where
\[
\begin{split}
r_n&=\frac{\ka_1}{\ka_2}[\int (e_{p,p}^T \G(u, I)e_{p,p})^2 du][\int (e_{p,p}^T \G(u, I)e_{p,p})^2 f_U(u)du][\int(e_{p,p}^T \G(u, I)e_{p,p})^4du]^{-1},\\
a_n&=\frac{\ka^2_1}{\ka_2}h^{-1}[\int (e_{p,p}^T \G(u, I)e_{p,p})^2 du]^2[\int (e_{p,p}^T \G(u, I)e_{p,p})^4 du]^{-1}.
\end{split}
\]
\end{theorem}
The limit null distribution of $T_2$ is actually the same as Remark 4.2 in \cite{fan2001}Fan et al., with the weight function $w(x)=1$. 
As a result, we could also use a weighted residual sum of squares in the test to offset the heteroscedastic influence. Let
\benn
\text{RSS}'_0 = \sum_{i=1}^k (\h a_p(\bar U_{i\cdot})-\h c)^2 w(\bar U_{i\cdot}), \quad \text{RSS}'_1 = \sum_{i=1}^k (\h a_p(\bar U_{i\cdot}) - \ti m_h(\bar U_{i\cdot}))^2 w(\bar U_{i\cdot})
\eenn
where $w(u)=[(e_{p,p}^T \G(u, I)e_{p,p})^2 \si^2]^{-1}$, then
\benn
T_2'= \frac{n}{2I}\log\frac{\text{RSS}'_0}{\text{RSS}'_1}.
\eenn
By Remark 4.2 in \cite{fan2001}, we know that $r'_n T'_2\overset{a}{\sim}\chi^2_{a'_n}$
with $r'_n=\frac{\ka_1}{\ka_2}$ and $a'_n=\frac{\ka^2_1}{\ka_2}h^{-1}|\Omega|$. Here $\Omega$ is the support of $U$ and $|\Omega|$ stands for the range of $U$. When weighted residual sum of squares are used, the asymptotic result is the same with that of GLR test directly applied on the original varying coefficient model. The difference is that our proposal saves a lot of computation. If we directly use GLR test for the original varying coefficient model, we have to estimate other $p-1$ functional coefficients under both null and alternative hypothesis. 

Next we consider the asymptotic distribution of $T_3$.
\begin{theorem}\label{thm:GLR2}
Suppose the assumptions (a1), (X), ($\ep$1), ($\ep$2), (U) and (K) hold and $I$ is a given positive integer. Then under $\cH_0$,
\[
\frac{2(I-p)}{I}\si_3^{-1}(T_3-\frac{n}{2(I-p)})\to N(0,1),
\]
where $\si_3^2=\Psi_n(\mu_4/\si^4-3)+2n/I$.
Furthermore, if $\ep$ follows a mesokurtic distribution(say, normal distribution), then
\[
\frac{2(I-p)}{I}T_3\overset{a}{\sim}\chi^2_{n/I}.
\]
\end{theorem}
When $\ep$ is distributed with a normal distribution, the null distribution of $T_3$ is quite simple. The underlying $\chi^2$ distribution is only related to the group size $I$, the covariates dimension $p$ and the sample size $n$. The Wilk's phenomenon is valid. Unlike $T_1$ and $T_2$, the estimation for the asymptotic mean and variance is not needed. This is a great merit of $T_3$. 

\section{Numerical Studies}\label{sec4}

\subsection{A real data example}\label{sec41}
In this section, we apply the proposed method to an environmental data set, which is also analyzed by \cite{fan1999}.
The data set records daily measurements of air pollutants and other environmental factors in Hong Kong from January 1, 1994 to December 31, 1995.
Here we want to study the association between the air pollutants level and the number of hospital admissions for circulation and respiration problem.
The air pollutants we considered are Sulphur Diocide, Nitrogen Dioxide and respirable suspended particulate, denoted as $X_2$, $X_3$ and $X_4$.
All are measured in $\mu g/m^3$.
The respond variable $Y$ represents the number of daily hospital admissions and $U=t=$time. Also we will include an intercept term $X_1=1$.

\begin{figure}
\includegraphics[width=\textwidth]{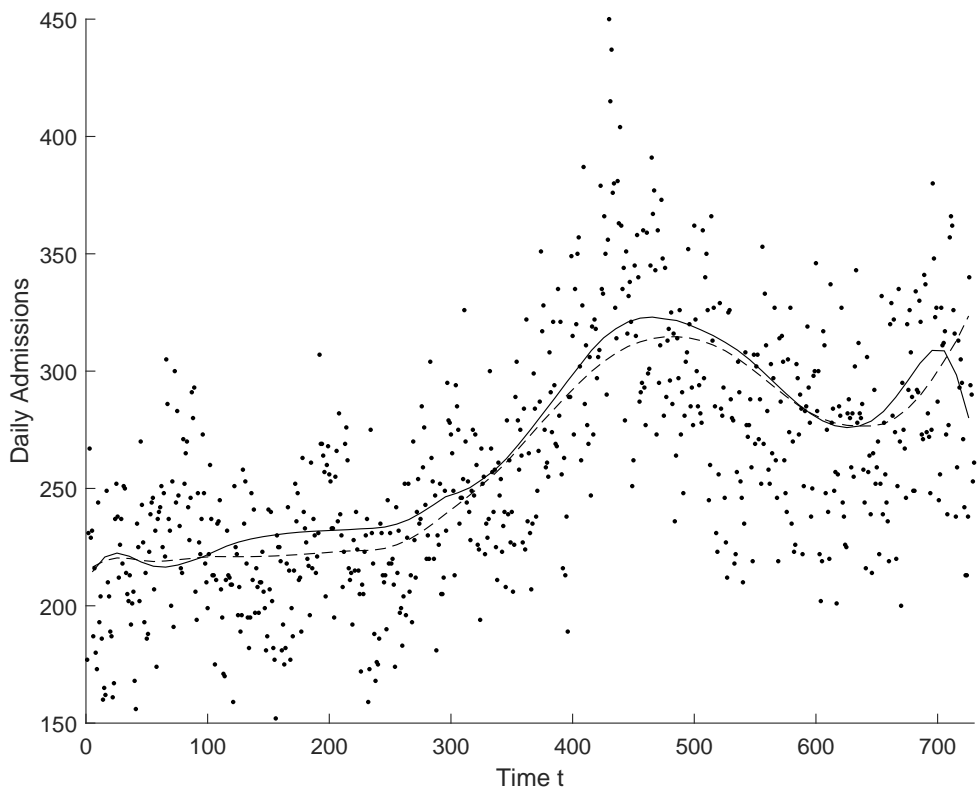}
\caption{Scatter of daily hospital admissions and expected curve when pollutant levels are set at averages. Solid line: full model. Dashed: deduced model.}
\label{fig:realdata}
\end{figure}

\begin{figure}
\includegraphics[width=\textwidth]{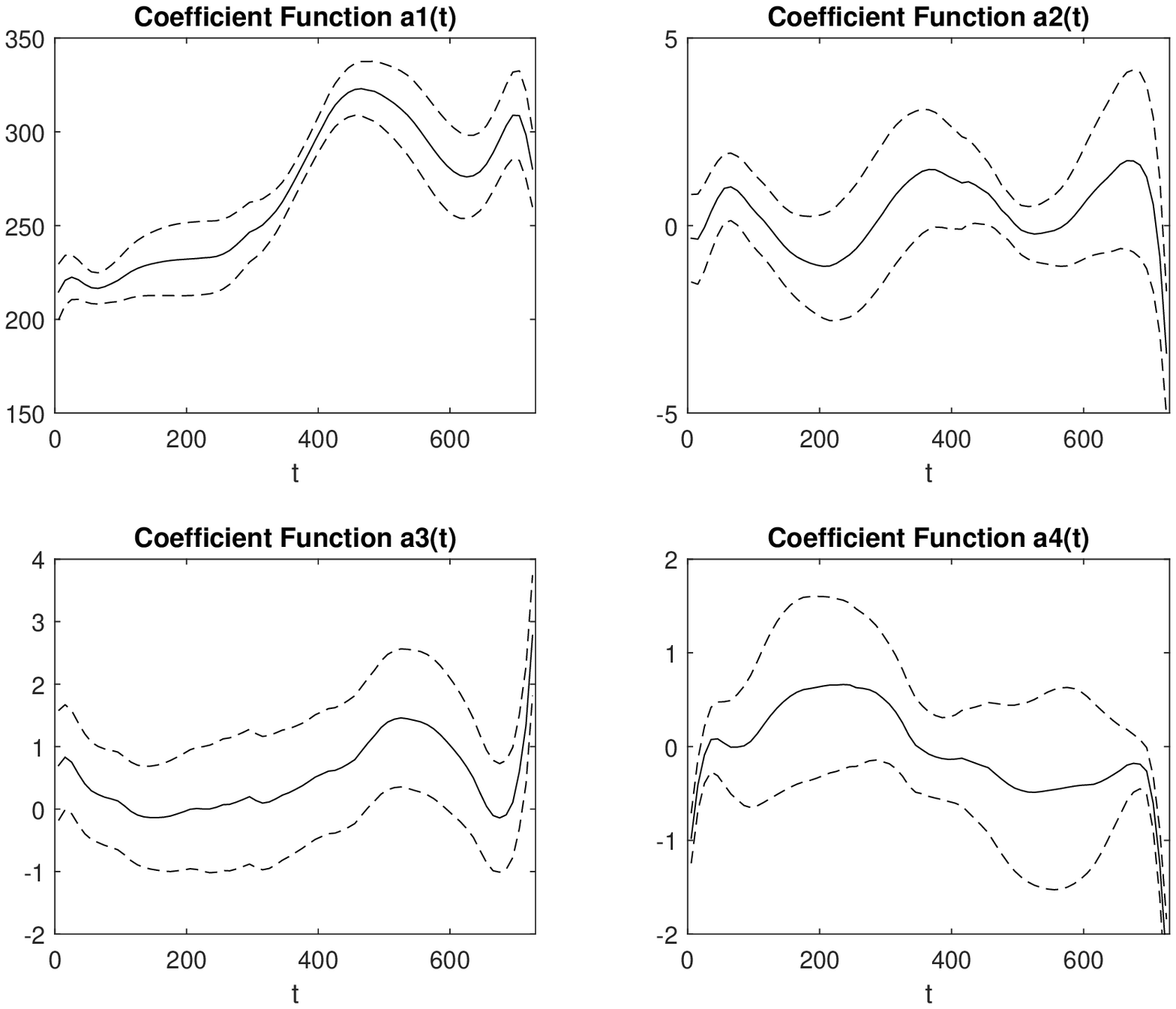}
\caption{The estimated coefficient functions with pointwise 95\% confidence intervals for the full model. }
\label{fig:4coefficients}
\end{figure}

Figure \ref{fig:realdata} shows the scatter plot of the daily number of hospital admissions for circulation and respiration against time $t$.
From this figure, one can see a clear increasing trend and some possible seasonal circular waves.
We center $X_2, X_3$ and $X_4$ and propose the following model to fit the data
\[
Y=a_1(t)+a_2(t)X_2+a_3(t)X_3+a_4(t)X_4+\ep.
\]
In this application, we choose $I=10$ and take $h$ to be $30\%$ of the interval length.
The estimated coefficient functions and their pointwise $95\%$ confidence bonds were shown in Figure \ref{fig:4coefficients}.
The confidence bonds are calculated directly from Theorem \ref{thm:vary1} with residual variance estimated by the method proposed in \cite{zhao2018}.
From Figure \ref{fig:4coefficients},  we can find that there is time effect on at least one coefficient.
In addition, the solid line in Figure \ref{fig:realdata} shows how the expected number of hospital admissions change over time when the pollutants levels are at their averages.
Now the increase in the Year 1995 and the seasonal effect are more obvious.

\begin{figure}
\includegraphics[width=\textwidth]{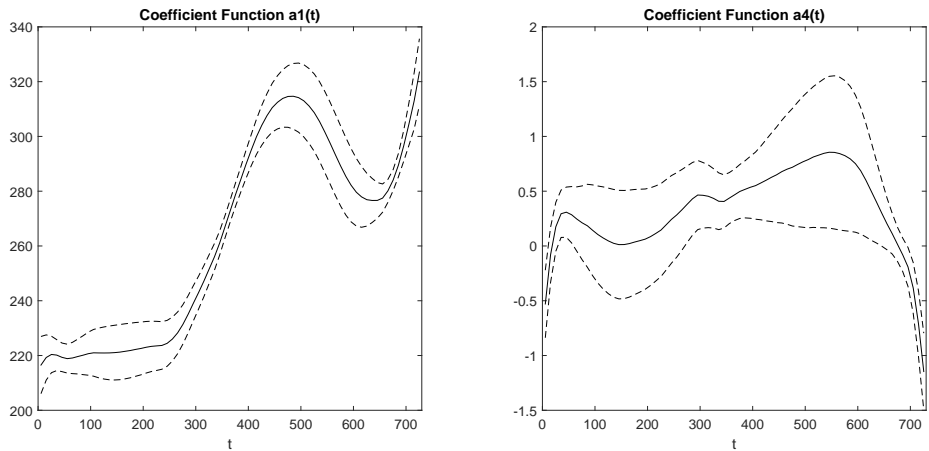}
\caption{The estimated coefficient functions for deduced model.}
\label{fig:2coefficients}
\end{figure}

Now we apply our test $T_3$-test to check whether the coefficients are really time varying or even significant.
Table \ref{tab:p-value} shows the p-values.
According to the p-values in Table \ref{tab:p-value}, we cannot reject the hypotheses $a_2(t)=0$ and $a_3(t)=0$. 
This result is different from that in \cite{fan1999}.
We remove the covariates $X_2, X_3$ and proposed a deduced model
\[
Y=a_1(t)+a_4(t)X_4+\ep
\]
Then we get the estimated coefficient functions and plot them in Figure \ref{fig:2coefficients}.
Compared with the coefficient functions in Figure \ref{fig:4coefficients}, the varying extent of the coefficients in the deduced model is more strong.
We also plot the expected number of hospital admissions under this deduced model.
It is shown in dashed in Figure \ref{fig:realdata}.
The overall trends of the two expected curves are alike and main differences appear at boundaries. 
In all, the daily hospital admissions for respiratory and circulatory shows an overall increasing trend and some seasonal patterns.

\begin{table}
\centering
\caption{The p-values for testing whether a coefficient function is zero(or a constant)}\label{tab:p-value}
\begin{tabular}{|l|l|l|l|l|}
\hline
& $a_1(t) $    & $a_2(t) $   & $a_3(t)$     & $a_4(u)$ \\
\hline
    $H_0: a_j(\cdot)=0$     & 0.0000 & 0.0832 & 0.0681 & 0.0460 \\
    $H_0: a_j(\cdot)=c$     & 0.0000 & 0.1100 & 0.0847 & 0.0482 \\
\hline
\end{tabular}
\end{table}

\subsection{Simulation for varying coefficient model}\label{sec42}

To investigate the performance of the proposed estimator (\ref{Step2Est}), we consider the following three examples:
\begin{align*}
\text{Example} \ 1.\hspace{0.2cm} Y=&\sin(60U)X_1+4U(1-U)X_2 + \sigma\ep.\\
\text{Example} \ 2.\hspace{0.2cm} Y=&\sin(6\pi U)X_1+\sin(2\pi U)X_2 + \sigma\ep.\\
\text{Example} \ 3.\hspace{0.2cm} Y=&\sin(8\pi(U-0.5))X_1\\
 &+\{3.5[\exp(-(4U-1)^2)+\exp(-(4U-3)^2)]-1.5\}X_2 + \sigma\ep
\end{align*}
where $U$ is uniformly distributed on $[0,1]$, $\ep$, $X_1$ and $X_2$ are generated from standard normal. 
Moreover, $Cov(X_1, X_2) = 2^{-1/2}$ and $\ep, U$ and $(X_1,X_2)^T$ are independent.
To make signal-to-noise ratio be about 5:1, $\si$ is chosen as
\[
\si^2=0.2\text{Var}[m(U,X_1,X_2)] \hspace{0.2cm} \text{with}\hspace{0.2cm} m(U,X_1,X_2)=\text{E}[Y|U,X_1,X_2].
\]
These examples were also used in \cite{fan1999} to study the performance of the one-step estimator and the two-step estimator.
For each example, the objective functional coefficient is $a_2$ and 100 replications are conducted with sample size $n=500$.
Mean integrated squared errors (MISE) are recorded to evaluate the performance of the estimators.
We consider the one-step estimator \citep{hoover1998} and the two-step estimator \citep{fan1999} as competitors.

\begin{figure}
\includegraphics[width=\textwidth]{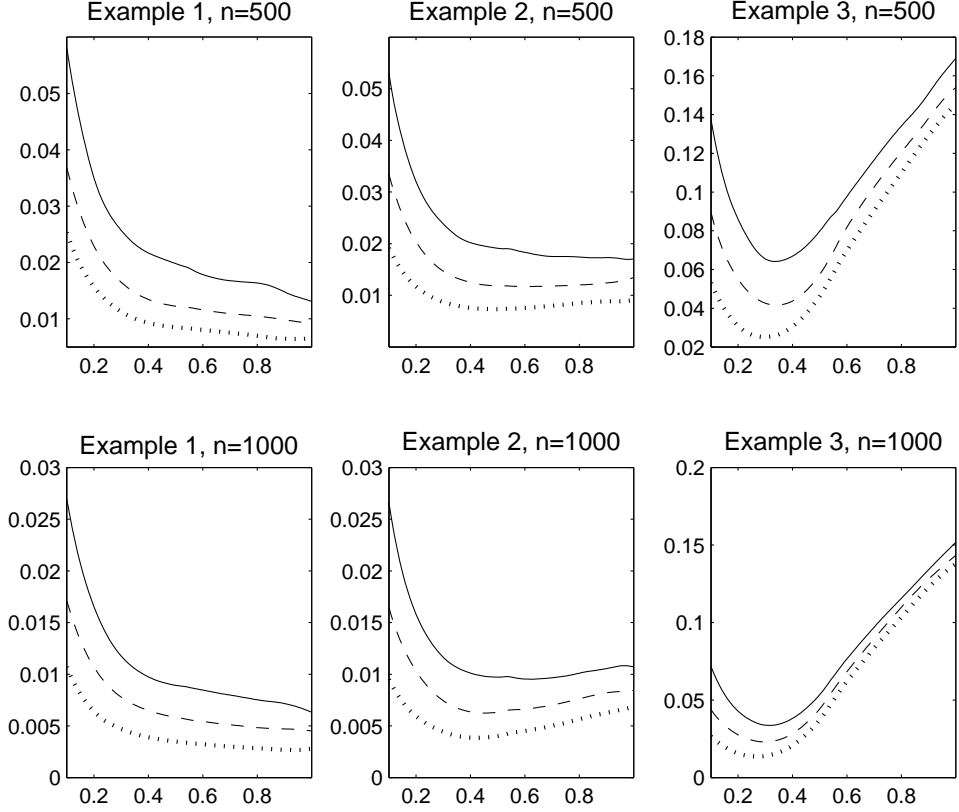}
\caption{MISE as a function of bandwidth. Solid curve: $I=4$; dashed curve: $I=5$; dotted curve: $I=10$. }
\label{fig:mise}
\end{figure}


In Figure~\ref{fig:mise}, we plot the MISE curve against bandwidth $h$ for each example when sample size $n=500$ and $n=1000$.
We can find that as $n$ increases, the estimation results become better.
A larger $I$ leads to a smaller asymptotic variance and have no influence on bias, so that the MISE becomes smaller.
However, one can notice that the improvement of $I=5$ from $I=4$ is almost the same with that of $I=10$ from $I=5$.
The marginal effect is decreasing quickly.
Therefore $I=10$ can already give a good estimation, though theoretically a large $I$ may be preferred.
One can also notice that the trends for different $I$ are similar.
This indicates that $I$ and $h$ in the smoothing step are independent. 
Thus it should not bother a lot to choose the group size $I$.

Next we compare the performance of the local average estimator, the one-step estimator and the two-step estimator.
The parameter $I$ is $10$.
The bandwidth $h$ is taken to be $0.2, 0.4, 0.6, 0.8, 1.0.$
Table~\ref{tab:MISE} reports the MISE of the three estimators.
In these cases, the MISE values of the local average estimator are always smaller than those of the one-step approach, which implies that the proposed method performs better than the one-step estimator.
On the other hand, the local average estimator is comparable to the two-step estimator.
This is consistent to Theorem~\ref{thm:vary1} and Theorem~\ref{thm:vary2}.

\begin{table}
\centering
\caption{The MISE of local average estimator, one-step estimator and two-step estimator.}\label{tab:MISE}
\begin{tabular}{|l|l|l|l|l|l|}
\hline
Example~1    & $h=0.2$& $h=0.4$& $h=0.6$& $h=0.8$ & $h=1.0$\\
\hline
local average& 0.0096 &	0.0104 & 0.0079	& 0.0096 & 0.0063 \\
one step     & 0.0285 & 0.0240 & 0.0151 & 0.0103 & 0.0112 \\
two step     & 0.0111 & 0.0112 & 0.0120 & 0.0076 & 0.0062 \\
\hline
Example~2        & $h=0.2$& $h=0.4$& $h=0.6$& $h=0.8$ & $h=1.0$\\
    \hline
local average& 0.0142 & 0.0106 & 0.0094 & 0.0089 & 0.0095 \\
one step     & 0.0900 & 0.0501 & 0.0460 & 0.0383 & 0.0399 \\
two step     & 0.0111 & 0.0087 & 0.0082 & 0.0077 & 0.0100 \\
\hline
Example~3        & $h=0.2$& $h=0.4$& $h=0.6$& $h=0.8$ & $h=1.0$\\
    \hline
local average& 0.0231 & 0.0384 & 0.0926	& 0.1382 & 0.1673 \\
one step     & 0.0808 & 0.0664 & 0.1093 & 0.1635 & 0.1976 \\
two step     & 0.0177 & 0.0344 & 0.1000 & 0.1351 & 0.1745 \\
\hline
\end{tabular}
\end{table}

An outstanding advantage of local average estimator is the computation simplicity. 
Table~\ref{tab:time_varying} shows the time spent of once implementation of local average estimator, two-step estimator and one-step estimator.
The time listed in the table are obtained by the function ``tic" ``toc" in MATLAB running with a dual 14-core cpu.
We can find the significant advantages of the local average estimator.
It is not difficult to find the reason.
In each estimator, most of the computations are involved in the weighted least squares process. 
For two step estimator and one step estimator, the weighted least squares process has to deal with a $n\times n $ matrix.
However, for local average estimator, the largest matrix size in weighted least squares process is $k\times k$.
Since $k=n/I$ and $I$, the matrix size of local average estimator is much smaller than that of the other two estimators in weighted least squares process.
In this way, the local average estimator saves a lot of computations.
It can be thought that in the ``average" step of our estimator, we have done some data mining to get a more corrected, ordered and simplified data set.
The ``average" step not only concentrates the information but also makes the disturbance abate.

\begin{table}
\centering
\caption{Typical time (in seconds) used by different estimators, I.}\label{tab:time_varying}
\begin{tabular}{|l|l|l|l|}
\hline
         &Example 1 &Example 2 & Example 3 \\
\hline
   local average & 0.21  & 0.20  & 0.20  \\
   two step      & 1.37  & 1.32  & 1.22  \\
   one step      & 1.41  & 1.39  & 1.48  \\
\hline
\end{tabular}
\end{table}

\subsection{Simulation for semivarying coefficient model}\label{sec43}

We consider the following semivarying coefficient models:
\begin{align*}
\text{Example} \ 4.\hspace{0.2cm} Y=&\sin(2\pi U)X_1+\cos(2\pi U)X_2+X_3 + \si\ep.\\
\text{Example} \ 5.\hspace{0.2cm} Y=&\sin(2\pi U)X_1+\{3.5[\exp(-(4U-1)^2)\\
                                    &+\exp(-(4U-3)^2)]-1.5\}X_2+X_3 + \si\ep.\\
\text{Example} \ 6.\hspace{0.2cm} Y=&\sin(6\pi(U-0.5))X_1+\sin(2\pi U)X_2+X_3 + \si\ep.
\end{align*}
where $U\sim U(0,1)$ ,and $\ep$, $X_i$, $i=1,2,3$, follows standard normal.
The $\si$ in each example is selected so that the signal-to-noise ratio is $5:1$. Further, $U$, $X_1$, $X_2$, $X_3$ and $\ep$ are mutually independent.

In this study, the sample size $n=500$ and the replication time is 100. For the constant coefficients, the mean, the standard error and the mean squared error(MSE) of the estimators are reported in Table \ref{tab:semila_constant}.
Form Table \ref{tab:semila_constant}, we can find that our estimators in these examples are close to the true value 1.
For different $I$, it makes no particular difference on the mean while a larger $I$ gives a smaller standard deviation.
This phenomena is consistent with Theorem~\ref{thm:semi}, since $I$ only appears in the asymptotic variance.

\begin{table}
\centering
\caption{Simulation results of the constant coefficients.}\label{tab:semila_constant}
\begin{tabular}{|l|l|l|l|l|l|l|l|l|l|l|}
\hline
          &       & \multicolumn{3}{l|}{Example 4} & \multicolumn{3}{l|}{Example 5} & \multicolumn{3}{l|}{Example 6} \\
\hline
    n     &    I  & mean   & std    & mse    & mean   & std    & mse    & mean   & std    & mse \\
\hline
    500   & 4     & 0.9992 & 0.0396 & 0.0016 & 0.9997 & 0.0558 & 0.0031 & 1.0001 & 0.0409 & 0.0017 \\
          & 5     & 1.0012 & 0.0364 & 0.0013 & 1.0019 & 0.0504 & 0.0025 & 1.0003 & 0.0350 & 0.0012 \\
          & 10    & 1.0000 & 0.0315 & 0.0010 & 1.0025 & 0.0448 & 0.0020 & 1.0011 & 0.0319 & 0.0010 \\
\hline
\end{tabular}
\end{table}

Next we compare the proposed estimator $\h \mbfb_{\text{LA}}$ in (\ref{Step2EstSemi}) with some existing estimators.
We consider \cite{zhang2002}'s estimator($\h \mbfb_{\text{Z}}$), \cite{fan2005}'s estimator($\h \mbfb_{\text{F}}$) and \cite{xia2004}'s estimator($\h \mbfb_{\text{X}}$) as competitors.
Another 100 replicates with sample $n=500$ of each example are generated and we use different methods to estimate the constant coefficient $b=1$.
Table~\ref{tab:semila_compare} reports the mean, the standard deviation and the MSE of these methods.
All the means are close to the true value.
The difference is less than 0.001, witch is a quite small error.
The standard deviation of local average estimator is the largest.
So the mse of the local average estimator is larger than other's.
We should have expected this result since Theorem~\ref{thm:semi} has already implied the inefficiency of local average estimator.

\begin{table}
\centering
\caption{Table captions should be placed above the tables.}\label{tab:semila_compare}
\begin{tabular}{|l|l|l|l|l|}
\hline
\multicolumn{5}{|l|}{Example 4} \\
\hline
      & $\h \mbfb_{\text{LA}}$ & $\h \mbfb_{\text{Z}}$ & $\h \mbfb_{\text{F}}$ & $\h \mbfb_{\text{X}}$ \\
 mean & 0.9997 & 0.9993 & 0.9995 & 0.9993 \\
 std  & 0.0313 & 0.0292 & 0.0287 & 0.0292 \\
 MSE  & 0.0010 & 0.0009 & 0.0008 & 0.0009 \\
\hline
\multicolumn{5}{|l|}{Example 5} \\
\hline
     & $\h \mbfb_{\text{LA}}$ & $\h \mbfb_{\text{Z}}$ & $\h \mbfb_{\text{F}}$ & $\h \mbfb_{\text{X}}$ \\
mean & 0.9997 & 0.9990 & 0.9993 & 0.9990 \\
 std & 0.0429 & 0.0400 & 0.0392 & 0.0398 \\
 MSE & 0.0018 & 0.0016 & 0.0015 & 0.0016 \\
\hline
 \multicolumn{5}{|l|}{Example 6} \\
\hline
     & $\h \mbfb_{\text{LA}}$ & $\h \mbfb_{\text{Z}}$ & $\h \mbfb_{\text{F}}$ & $\h \mbfb_{\text{X}}$ \\
mean & 0.9998 & 0.9992 & 0.9995 & 0.9992 \\
 std & 0.0316 & 0.0294 & 0.0289 & 0.0293 \\
MSE  & 0.0010 & 0.0009 & 0.0008 & 0.0009 \\
\hline
\end{tabular}
\end{table}

Here we still want to discuss the computation simplicity, which is the significant advantage of the proposed method.
Table~\ref{tab:time_semi} shows the time used by the above mentioned estimators.
The same with Table~\ref{tab:time_varying}, we use the function ``tic" ``toc" in MATLAB to do the timing and run those codes with a dual 14-core cpu.
Here we can see a huge advantage of the local average estimator.
The time spent by other estimators are tens of that spend by local average estimator.

\begin{table}
\centering
\caption{Typical time (in seconds) used by different estimators, II}\label{tab:time_semi}
\begin{tabular}{|l|l|l|l|}
\hline
           &  Example 4 &  Example 5 &  Example 6 \\
\hline
 $\h \mbfb_{\text{LA}}$  &      0.01  &      0.01  &      0.01  \\
 $\h \mbfb_{\text{Z}}$   &      0.34  &      0.32  &      0.29  \\
 $\h \mbfb_{\text{F}}$   &      0.45  &      0.47  &      0.39  \\
 $\h \mbfb_{\text{X}}$   &     14.47  &     13.12  &     13.83  \\
\hline
\end{tabular}
\end{table}

From all these simulations, we can conclude that the proposed estimator $\h \mbfb_{\text{LA}}$ can give a good estimation and dramatically reduce the computation burden.
Though it is not asymptotic efficient, local average estimator can be a good primary estimator or pilot estimator.

\subsection{Size and power study for the proposed tests}\label{sec44}

In this section, we investigate the finite-sample performance of the proposed tests. Consider the model
\benn
Y=a_1(U)X_1+a_2(U)X_2+\ep,
\eenn
where $U\sim U(0,1)$ ,and $\ep$, $X_i$, $i=1,2,3$, follow standard normal.
The $\si$ in each example is selected so that the signal-to-noise ratio is $5:1$. Further, $U$, $X_1$, $X_2$, $X_3$ and $\ep$ are mutually independent.
The problem of interest is to test:
\benn
\cH_0:  a_2(u)=c, \quad \text{vs} \quad \cH_1: a_2(u)\neq c
\eenn
First we consider the null model with $a_1(u)=\sin(60u)$ and $a_2(u)=1$.
The replication time is 1000 and the significance level $\al$ is 0.05.
Then we calculated the empirical size of the three proposed tests.
The bandwidth $h$ is taken to be $n^{-2/5}$ and $n^{-1/5}$ for $T_1$ and $T_2$ respectively \citep{zheng1996,fan1996}.

\begin{table}
\centering
\caption{Proportion of rejections for null model with $T_1$ and $T_2$}\label{tab:T12size}
\begin{tabular}{|ll|lll|}
\hline
$T_1$&  $n$  &$I = 4$&$I = 5$&$ I = 10$  \\
    \hline
     & 400   & 0.037 & 0.036 & 0.039 \\
     & 800   & 0.040 & 0.036 & 0.038 \\
     & 1600  & 0.042 & 0.042 & 0.036 \\
\hline
$T_2$&  $n$  &$I = 4$&$I = 5$&$ I = 10$  \\
    \hline
     & 400   & 0.075 & 0.079 & 0.089 \\
     & 800   & 0.069 & 0.070 & 0.074 \\
     & 1600  & 0.067 & 0.058 & 0.066 \\
\hline
$T_3$&  $n$  &$I = 4$&$I = 5$&$ I = 10$  \\
    \hline
     & 400   & 0.118 & 0.078 & 0.057 \\
     & 800   & 0.084 & 0.075 & 0.049 \\
     & 1600  & 0.081 & 0.053 & 0.052 \\
\hline
\end{tabular}
\end{table}


We summarize the results of size study of $T_1$, $T_2$ and $T_3$ in Table \ref{tab:T12size}.
As can be seen, in most cases the test $T_1$ has size close to 0.05. When sample size $n$ becomes larger, the sizes tend to the asymptotic value.
What's more, when the sample size is large enough, the influence caused by the choice of group size $I$ seems slight.
For $T_2$, the sizes get closer to 0.05 as $n$ increases.
However we can not tell the difference among different bandwidths and the group sizes.
Notice that the sizes in the table are all larger than 0.05, though the convergent trend exits.
As to $T_3$, its performance is satisfactory. The sizes of $T_3$ converge to 0.05 rapidly as $n$ grows.
What's more, the test statistics with $I=10$ outperform those of the other two cases. This is consistent with the results in Theorem~\ref{thm:GLR2}. 
In order to have a more intuitional understanding about the asymptotic distribution of the test statistics under null hypothesis, we plot the empirical density functions of the three proposed test statistics under the null model.
In addition, all the test statistics are standardized so that we can compare the sample distributions with the standard normal distribution.
Group size $I$ are selected as 10 in all the three test statistics.

\begin{figure}
\includegraphics[width=\textwidth]{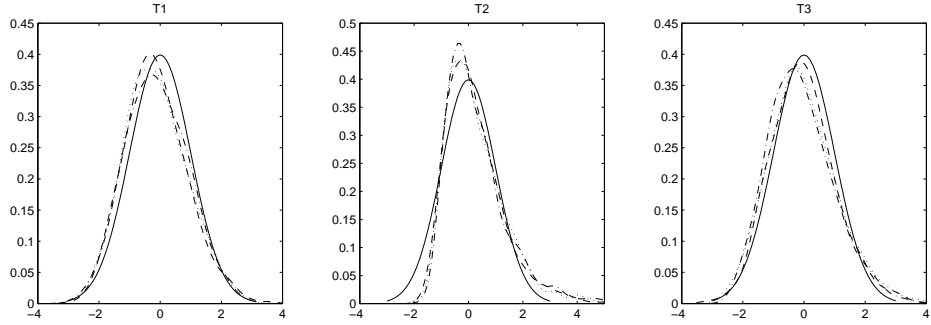}
\caption{Null distributions of  test statistics $T_1$, $T_2$ and $T_3$. Solid curve: standard normal; dotted curve: n=400; dash-dot curve: n=800; dashed curve: n=1600 . }
\label{fig:nulldistribution}
\end{figure}


It can be seen from Figure~\ref{fig:nulldistribution} that the sample distributions of standardized $T_1$ and $T_3$ have a similar bell shape as the standard normal distribution.
What's more, the sample distributions behave like the standard normal more as sample size $n$ gets larger.
For the test statistics $T_2$, there seems to present some discrepancy between the sample and the standard normal distribution.
Thought the sample distribution is quite close to the normal standard, we can still find a long right tail.
This may explain some of the facts that the size of test $T_2$ is usually larger that the significance level.

Next we conduct the power study of the proposed tests.
Take the following two families of alternative models as examples:
\begin{align*}
&\text{Example 7}.\quad a_1(u)=\sin(60u), \quad a_2(u)=a\cdot4u(1-u)+(1-a).\\
&\text{Example 8}.\quad a_1(u)=\sin(6\pi u), \quad a_2(u)=a\cdot \sin(2\pi u)+(1-a).
\end{align*}
with the parameter $a=0, 0.1,...,1$. Obviously, the null hypothesis holds when $a=0$.
Then the functional coefficient $a_2(u)$ gradually departs from the constant as $a$ arises to 1.

\begin{figure}
\includegraphics[width=\textwidth]{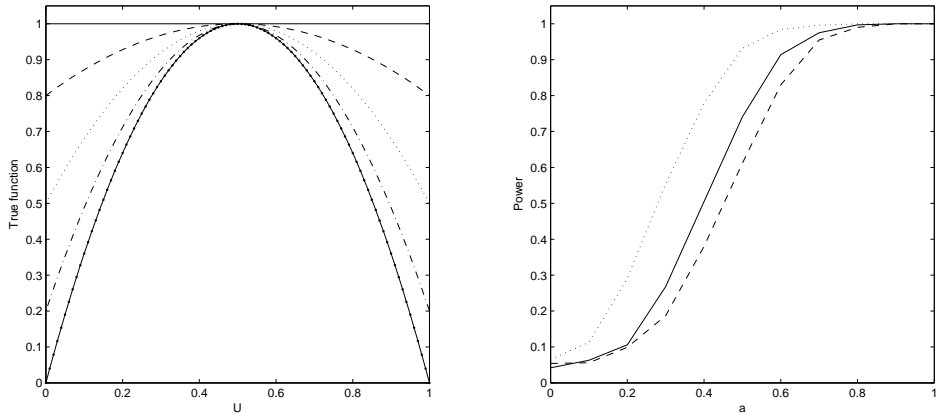}
\caption{Example 7: Left: True function when $a=0$(solid), $a=0.2$(dashed), $a=0.5$(dotted), $a=0.8$(dash-dotted), $a=1$(dotted-solid). Right: Power functions for the proposed tests under different alternatives. Solid curve: $T_1$; dotted curve: $T_2$; dashed curve: $T_3$.}
\label{fig:ex1power}
\end{figure}

\begin{figure}
\includegraphics[width=\textwidth]{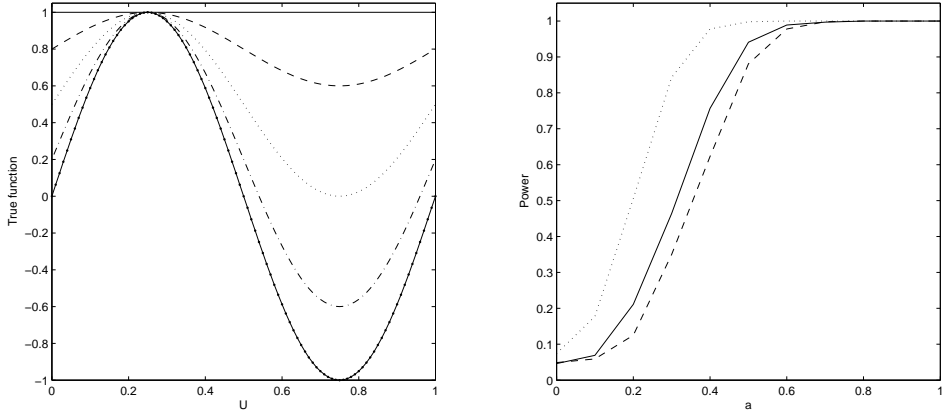}
\caption{Example 8: Left: True function when $a=0$(solid), $a=0.2$(dashed), $a=0.5$(dotted), $a=0.8$(dash-dotted), $a=1$(dotted-solid). Right: Power functions for the proposed tests under different alternatives. Solid curve: $T_1$; dotted curve: $T_2$; dashed curve: $T_3$.}
\label{fig:ex2power}
\end{figure}

Under these two families of alternative models, we compute the power functions of the three proposed tests.
The left panel of Figure~\ref{fig:ex1power} plots the true curve of the functional coefficient $a_2(u)$ in Example 7, ranging from the null hypothesis to the alternatives.
The right panel depicts the empirical power at 0.05 significance level.
It can be seen that all the three power functions increase to 1 rapidly, indicating the sensitivity for detecting the alternatives.
Figure \ref{fig:ex2power} shows the true functions of $a_2(\cdot)$ and the power functions of the tests at 0.05 significance level in Example~8.
As expected, the results reveal the proposed test statistics are powerful to detect the alternatives.

\section{Discussion}\label{sec5}

In this paper, we propose a fast inference procedure via local averaging to estimate functional coefficients of the varying coefficient model.
Furthermore, we extend it to the semivarying coefficient model. 
Both of the theoretical and simulation results show that the proposed estimators have good performance.
For the varying coefficient model, our estimator can easily deal with the different smoothness problem and reach an optimal convergence rage $n^{-8/9}$.
For the semivarying coefficient model, the proposed estimator for the constant part is asymptotically unbiased and asymptotic normal, and can be written as a form of a projection-based estimator.
The most impressive contribution of our estimators is the computation simplicity.
With a ``over-parameterized" step, we concentrate the information and decrease the sample size.
For model checking problems, our proposed tests can focus on testing one coefficient function and leave out all smoothing procedures of estimating nuisance coefficient.
Thus, we dramatically improve the efficiency.

As we mentioned before, the local average estimator (\ref{primary}) is a good base for further inference.
We have shown that how to build estimation and testing procedures by local averaging. 
Another important application is variable selection, which can significantly enhances the prediction accuracy of the fitted model if the underlying model has a sparse representation.
We take the semivarying coefficient model as example.
The works focus on variable selection for semivarying coefficient models seems scant.
\cite{li2008} proposed a variable section procedure by using nonconcave penalized likelihood.
They replaced $a_i(u)$, $i=1,\ldots,d$ by their local linear estimates and used SCAD \citep{fan2001scad} to obtain sparse estimate of $\mathbf{b}$.
\cite{liang2009} considered variable selection for partially linear models when the covariates are measured with additive errors.
In \cite{kai2011}, they propose adaptive penalization methods for semivarying coefficient models and prove that the methods possess the oracle property.
The computation cost of these methods are even severe due to the tuning procedure of the penalty parameter.
However the local average estimator can provide a good solution to save computation.
According to (\ref{PLS}) and \cite{zou2006}, we can propose a penalized error function
\benr\label{eq31}
 \text{E}_{\la_n}(\mbfb) = \text{E}(\mbfb) + \la_n \sum_{i=1}^p w_i |b_i| = \|\mathds{P} (\mathds{Y} - \mathds{Z} \mbfb)\|^2 + \la_n \sum_{i=1}^p w_i |b_i|,
\eenr
where $w_i = 1/|\h b_i|^\al$ are data-dependent weights, $\h b_i$ is the $i$-th element of  $\h{\mbfb}$ in (\ref{estb}), $\al$ is a  scaler and $\la_n$ is a sequence of constants.
The adaptive Lasso for semivarying coefficient model via local averaging is given by $\h{\mbfb}(\la_n) = (\h b_1(\la_n), \ldots, \h b_p(\la_n))^\T = \arg\min_{\mbfb} \text{E}_{\la_n}(\mbfb).$
We denote this estimator as adaptive LA-Lasso.
Note that $\text{E}(\mbfb)$ is the square error function of a linear model. 
According to \cite{zou2006}, the oracle property and the asymptotic normality of the adaptive LA-Lasso can be easily derived from those of the adaptive Lasso under linear model.

\newpage

\appendix
\section*{Appendix}

\begin{subsection}{Proof of Lemma~\ref{th:localaverage}}

In this section, we shall prove {\bf Lemma}~\ref{th:localaverage} from Section~\ref{sec21}:

{\bf Proof:} Note that the primary point estimators $\mbfa = (\mbfa_1^T, \mbfa_2^T, \cdots, \mbfa_k^T)^T$ is a result of ordinary least square from $k$ independent linear regressions. 
The components $\mbfa_i$, $i=1,\ldots,k$ are actually calculated separately. 
So without losing generality, we will discuss $\mbfa_i$ only.  
Rewrite $\hat{\mbfa}_i$ as
\benrr
\hat{\mbfa}_i = (\hat{a}_1(\bar{U}_{i\cdot}),\hat{a}_2(\bar{U}_{i\cdot}),\cdots,\hat{a}_p(\bar{U}_{i\cdot}))^T = (\bX_i^{*T} \bX_i^*)^{-1}\bX_i^{*T} \bY^*_i,
\eenrr
where 
\benrr
\bY_i = (Y_{(iI - I +1)}, Y_{(iI - I +2)}, \ldots, Y_{(iI)} )^T, \quad Y_{(iI - I +j)} = \bX^T_{(iI - I +j)}\mbfa(U_{(iI - I +j)}) + \ep_{(iI - I +j)}.
\eenrr
Add and subtract $\bX^T_{(iI - I +j)}\mbfa(\bar{U}_{i\cdot})$ into $Y_{(iI - I +j)}$, we have
\benrr
\hat{\mbfa}_i  =  \mbfa(\bar{U}_{i\cdot}) + (\bX_i^{*T}\bX_i^*)^{-1}\bX_i^{*T} \bmep_i^* + (\bX_i^{*T}\bX_i^*)^{-1} \bX_i^{*T} \Delta_i,
\eenrr
where $\bmep^*_i = (\ep_{(iI - I +1)}, \ep_{(iI - I +2)}, \ldots, \ep_{(iI)} )^T$ and  $\Delta_i = (\Delta_{i1}, \Delta_{i2},\ldots, \Delta_{iI})$ with
\benrr
 \Delta_{ij} = \bX_{(iI - I +j)}^T(\mbfa(U_{(iI - I +j)})-\mbfa(\bar{U}_{i\cdot})), \,\, \text{for}\,\, j =1, \ldots, I.
\eenrr
Now we shall prove that $|a_l(U_{(iI - I +j)})-a_l(\bar U_{i\cdot})|=O_p(\ln n/n)$ for any $l=1,\ldots,p$, $i =1,\ldots, k$ and $j =1, \ldots,I$.
Let $F_U(\cdot)$ be the cumulative distribution of $U$, i.e., $F_U'(u)=f_U(u)$. By mean value theorem, for a $\xi_{ij}$ is between $U_{(iI - I +j)} $ and $\bar{U}_{i\cdot}$,
\benrr
|a_l(U_{(iI - I +j)}) -a _l(\bar{U}_{i\cdot})|&= &|a_l'(\xi_{ij})||U_{(iI - I +j)} - \bar{U}_{i\cdot}|\\
&\leqslant&|a_l'(\xi_{ij})|\cdot\frac{I-1}{2}\max_{1\leq i \leq n}|U_{(i+1)}-U_{(i)}|.
\eenrr
Let $\tau=F_U(U)$, so we can regard $\tau$ as a uniformly distributed variable in the interval $[0,1]$. We denote two consecutive order statistics by $U_{(i+1)}, U_{(i)}$, and $\tau_{(i+1)},\tau_{(i)}$ are the corresponding uniformly distributed variables. By Assumption (U) and mean value theorem, we have
\benrr
&&\max_{1\leq i \leq n}|U_{(i+1)}-U_{(i)}|  = \max_{1\leq i \leq n}|F_U^{-1}(\tau_{(i+1)})-F_U^{-1}(\tau_{(i)})| \\
& = &  \max_{1\leq i \leq n} (F^{-1})'(\eta_i)|\tau_{(i+1)}-\tau_{(i)}| =  \max_{1\leq i \leq n} \frac{1}{f(u_{\eta_i})}|\tau_{(i+1)}-\tau_{(i)}|\\
& \leq & \frac{1}{\delta} \max_{1\leq i \leq n}|\tau_{(i+1)}-\tau_{(i)}| =   \frac{1}{\delta}  O_p(\frac{\ln n}{n})
\eenrr{}
where $\eta_i$ is between $\tau_{(i+1)}$ and $\tau_{(i)}$, $u_{\eta_i} = F^{-1}(\eta_i)$. The last equation holds by the Theorem 3.1 of \cite{holst1980}.
 Therefore, $\|\Delta_i\| \leq O_p(\ln n /n)$ and
 \benrr
 \hat{\mbfa}_i=\mbfa(\bar U_{i\cdot}) + (\bX_i^{*T} \bX_i^*)^{-1} \bX_i^{*T} \bmep_i^*+ O_p(\frac{\ln n}{n}).
 \eenrr
Further, one can see that, for any given $\bar{U}_{i\cdot}$,
\benrr
\text{E}(\hat{\mbfa}_i)=\mbfa(\bar{U}_{i\cdot}) + O(\frac{\ln n}{n}), \,\, \text{and}\,\,\, \text{Var}(\hat{\mbfa}_i) = \text{E}[(\bX_i^{*T}\bX_i^*)^{-1}|\bar{U}_{i\cdot}]\si^2 = \Gamma(\bar{U}_{i\cdot})\si^2.
\eenrr
What's more, since the ordering is no longer needed in the following smoothing step, we can naively consider the first phase estimators $(\bar{U}_{i\cdot},\hat{\mbfa}_i), 1=1,\ldots,k$ are independent and identically distributed. The proofs of {\bf Theorem}~\ref{thm:vary1} and {\bf Theore}~\ref{thm:vary2} are similar and simple. Combine {\bf Lemma}~\ref{th:localaverage} and {\bf Theorem}~3.1 in \cite{fan1996}, we will get the final results.

\end{subsection}

\begin{subsection}{Proof of Theorem~\ref{thm:semi}}

Note that $\bY_i^* = (Y_{(iI - I +1)}, Y_{(iI - I +2)}, \ldots, Y_{(iI)})^T$ where
\benrr
Y_{(iI - I +j)} = \bX^T_{(iI - I +j)}\mbfa(U_{(iI - I +j)}) + \bZ^T_{(iI - I +j)} \mbfb +\ep_{(iI - I +j)}.
\eenrr
Add and subtract $\bX^T_{(iI - I +j)}\mbfa(\bar{U}_{i\cdot})$ into $Y_{(iI - I +j)}$, we have
\benrr
Y_{(iI - I +j)} = \Delta_{ij} + \bX^T_{(iI - I +j)}\mbfa(U_{i\cdot})  + \bZ^T_{(iI - I +j)} \mbfb +\ep_{(iI - I +j)},
\eenrr
where  $\Delta_{ij} = \bX_{(iI - I +j)}^T(\mbfa(U_{(iI - I +j)})-\mbfa(\bar U_{i\cdot}))$.
Similar to the augment in the proof of {\bf Lemma}~\ref{th:localaverage}, we know $|\Delta_{ij} | \leq O_p(\ln n/n)$. Thus,
\benrr
\bY^*_i = O_p(\frac{\ln n}{n})+ \bX_i^{*T}\mbfa(\bar{U}_{i\cdot})+ \bZ_i^{*T} \mbfb + \bmep^*_i.
\eenrr
Then plug  the above equation into $\hat{\mbfb}$, we have
\benr\label{S1}
\hat{\mbfb} - \mbfb &=& (\mbf{0}_{1\times kp}, \mbf{1}_{1\times q}) (\bmPhi^T \bmPhi)^{-1} \bmPhi^T O_p(\frac{\ln n}{n}) + (\mbf{0}_{1\times kp}, \mbf{1}_{1\times q}) (\bmPhi^T \bmPhi)^{-1}\bmPhi^T \bmep
\eenr
Then we have,
\benr\label{S2}
\sqrt{n}(\hat{\mbfb} - \mbfb) &=& (\mbf{0}_{1\times kp}, \mbf{1}_{1\times q}) (\bmPhi^T \bmPhi)^{-1} \bmPhi^T \mbf{O}_p(\frac{\ln n}{\sqrt{n}}) + \sqrt{n}(\mbf{0}_{1\times kp}, \mbf{1}_{1\times q}) (\bmPhi^T \bmPhi)^{-1} \bmPhi^T \boldsymbol{\epsilon} \nn \\
&=&\sqrt{n}(\mbf{0}_{1\times kp}, \mbf{1}_{1\times q}) (\bmPhi^T\bmPhi)^{-1}\bmPhi^T \bmep + o_p(1)
\eenr
To proceed further, we denote
\benrr
\mbf{A} = diag(\bX^{*T}_1 \bX_1^*,\dots, \bX^{*T}_k \bX_k^*), \quad \mbf{B} = (\bZ^{*T}_1 \bX_1^*, \dots, \bZ_k^{*T}\bX_k^*)^T,\quad \mbf{C} =\ sum_{i=1}^k \bZ_i^{*T} \bZ_i^*.
\eenrr
Then we can write $\bmPhi^T\bmPhi =\begin{pmatrix}
 \mbf{A}  & \mbf{B} \\
 \mbf{B}^T & \mbf{C}\\
\end{pmatrix}$ and
\benrr
(\bmPhi^T\bmPhi)^{-1} = \begin{pmatrix}
                   \mbf{A}^{-1} + \mbf{A}^{-1} \mbf{B}(\mbf{C}-\mbf{B}^T\mbf{A}^{-1}\mbf{B})^{-1}\mbf{B}^T\mbf{A}^{-1} & -\mbf{A}^{-1}\mbf{B}(\mbf{C}-\mbf{B}^T\mbf{A}^{-1}\mbf{B})^{-1}\\
                   -(\mbf{C}-\mbf{B}^T\mbf{A}^{-1}\mbf{B})^{-1}\mbf{B}^T\mbf{A}^{-1}   & (\mbf{C}-\mbf{B}^T\mbf{A}^{-1}\mbf{B})^{-1}
\end{pmatrix}.
\eenrr
Plug the equation of$(\bmPhi^T \bmPhi )^{-1}$ into (\ref{S2}),  we can have
\benrr
\sqrt{n}(\mbf{0}_{1\times kp}, \mbf{1}_{1\times q})  (\bmPhi^T \bmPhi)^{-1} \bmPhi^T \bmep = R_1^{-1}R_2
\eenrr
where
\benrr
R_1 & = & \frac{1}{n}\sum_{i=1}^k \bZ_i^{*T} \bZ_i^* - \bZ^{*T}_i \bX_i^*(\bX^{*T}_i \bX^*_i)^{-1} \bX^{*T}_i \bZ_i \\
R_2 &=& \frac{1}{\sqrt{n}}\sum_{i=1}^k(\bZ_i^{*T}- \bZ^{*T}_i \bX^*_i (\bX^{*T}_i \bX^*_i)^{-1})\bmep^*_i.
\eenrr
First consider $R_1$. The expectation of $R_1$ is calculated as follow.
\benrr
\text{E}[R_1] &=& \text{E}\Big[\frac{1}{I}\sum_{j=I}^I \bZ_{(iI -I +j)} \bZ_{(iI -I +j)}^T- (\frac{1}{I}\sum_{j=I}^I \bZ_{(iI -I +j)} \bX_{(iI -I +j)}^T) \\
&& \times (\frac{1}{I} \sum_{j=I}^I \bX_{(iI -I +j)} \bX_{(iI -I +j)}^T)^{-1} (\frac{1}{I}\sum_{j=I}^I \bX_{(iI -I +j)} \bZ_{(iI -I +j)}^T)\Big] \\
&=& \text{E}\Big[\frac{1}{I}\sum_{j=I}^I\text{E}[ \bZ_{(iI -I +j)} \bZ_{(iI -I +j)}^T|U_{(iI -I +j)}]\Big] - \text{E}\Big[\text{E}\big[(\frac{1}{I} \sum_{j=I}^I \bZ_{(iI -I +j)} \bX_{(iI -I +j)}^T) \\
&& \times (\frac{1}{I}\sum_{j=I}^I \bX_{(iI -I +j)} \bX_{(iI -I +j)}^T)^{-1} (\frac{1}{I} \sum_{j=I}^I \bX_{(iI -I +j)} \bZ_{(iI -I +j)}^T)|U_{(iI -I +1)},...U_{(iI)}\big]\Big]\\
\eenrr
We can see that $\text{E}[R_1] = \Sigma$. Further, by law of large numbers, $R_1$ converges in probability to $\Sigma$ as $k\to \infty$.

Next we deal with the term $R_2$. Regardless of the ordering and given $\{(U_i, \bX_i, \bZ_i)\}$, $i=1,...,n$, $\bmep_i$ is independent of each other and has mean zero. Therefore, $R_2$ is asymptotically normal with mean zero. Then we only need to investigate the limit variance of $R_2$. Similar to the augment for the expectation of $R_1$, we know
\benrr
\text{Var}(R_2|\{U_i, \bX_i, \bZ_i\}) = \frac{\sigma^2}{n}\sum_{i=1}^k(\bZ_i^{*T} \bZ_i^* - \bZ^{*T}_i \bX_i^* (\bX^{*T}_i \bX_i^*)^{-1} \bX^{*T}_i \bZ_i^*) \rightarrow_p \sigma^2 \Sigma.
\eenrr
Therefore, by the Slutsky theorem,
\benn
\sqrt{n}(\hat{\mbfb} - \mbfb) \rightarrow N(\mbf{0}, \sigma^2\Sigma^{-1})
\eenn

\end{subsection}

\begin{subsection}{Proof of Theorem~\ref{thm:T1}}
By {\bf Lemma}~\ref{th:localaverage},
\benrr
\hat{a}_p(\bar{U}_{i\cdot})=a_p(\bar{U}_{i\cdot})+\eta_i + O_p(\frac{\ln n}{n}), \,\, i=1,\ldots,k
\eenrr
where the new error terms $\eta_i, i=1,...,k$ are independent and have zero mean.
It is easy to see that the bias term $O_p(\ln n/n)$ is asymptotically negligible after timing $kh^{1/2}$.
So we only consider $a_p(\bar{U}_{i\cdot})$ and $\eta_i$ in the following.
Under null hypothesis, $\hat{a}_p(\bar{U}_{i\cdot})=c+\eta_i$. Thus 
\benrr
\hat{c}=c+\frac{1}{k}\sum_{i=1}^k\eta_i,\quad \text{and} \quad \hat{a}_p(\bar{U}_{i\cdot})-\hat{c}=-\sum_{j\neq i}^k\frac{1}{k}\eta_j+\frac{k-1}{k}\eta_i. 
\eenrr
Then we can decompose $T_1$ into $S_1 + S_2 +S_3$ where
\benrr
S_1 &=& \frac{1}{k^3(k-1)} \sum_{i=1}^k \sum_{j\neq i}^k\sum_{s\neq i}^k \sum_{t\neq j}^k \frac{1}{h} K(\frac{\bar{U}_{i\cdot}-\bar{U}_{j\cdot}}{h})  \eta_s  \eta_t\\
S_2 &=& \frac{-2}{k^3} \sum_{i=1}^k \sum_{j\neq i}^k \sum_{s\neq i}^k \frac{1}{h}K(\frac{\bar{U}_{i\cdot}-\bar{U}_{j\cdot}}{h}) \eta_s \eta_j\\
S_3 &=& \frac{(k-1)}{k^3}\sum_{i=1}^k\sum_{j\neq i}^k\frac{1}{h}K(\frac{\bar{U}_{i\cdot}-\bar{U}_{j\cdot}}{h})\eta_i\eta_j.
\eenrr
It is easy to see that $\text{E}[S_1]=\text{E}[S_2]=O(1/k)$, and $\text{E}[S^2_1]=O(1/k^2)+O(1/(k^4h))$, $\text{E}[S^2_2]=O(1/k^2)+ O(1/(k^3h))$. 
Then by Chebyshev's inequality, we have $kh^{1/2}S_1=o_p(1)$ and $kh^{1/2}S_2=o_p(1)$.
Rewrite $S_3$ as $(\frac{k-1}{k})^2S_3^*$ where 
\benrr
S_3^*=\frac{1}{k(k-1)}\sum_{i=1}^k\sum_{j\neq i}^k\frac{1}{h}K(\frac{\bar{U}_{i\cdot}-\bar{U}_{j\cdot}}{h})\eta_i\eta_j.
\eenrr
By {\bf Lemma} 3.3 in \cite{zheng1996}, we know $kh^{1/2}S_3^* \Rightarrow N(0,\Sigma)$, where "$\Rightarrow$" stands for convergence in distribution and $\Sigma=2\int K^2(s)ds\cdot\int\{\text{E}[\eta^2|u]\}^2f(u)d(u)$.
By Slusky's theorem, $nh^{1/2}T_1\overset{d}{\to}N(0, \Sigma_1).$

Next we shall prove that $\Sigma_1$ can be consistently estimated by $\hat{\Sigma}_1$. Plug $\hat{a}_p(\bar{U}_{i\cdot})-\hat{c}=-\sum_{j\neq i}^k\frac{1}{k}\eta_j+\frac{k-1}{k}\eta_i$ into $\hat \Sigma_1$. Then the expectation of $\hat{\Sigma}_1$ can be decomposed into a sum of three terms:
 $\text{E}[S_4+S_5+S_6]$, where
\[
\begin{split}
S_4&=\frac{2I^2}{k(k-1)}\sum_{i=1}^k\sum_{j\neq i}^k\frac{1}{h}K^2(\frac{\bar{U}_{i\cdot}-\bar{U}_{j\cdot}}{h})(\sum_{s\neq i}^k\frac{1}{k}\eta_s)^2(\sum_{t\neq j}^k\frac{1}{k}\eta_t)^2,\\
S_5&=\frac{4I^2}{k(k-1)}\sum_{i=1}^k\sum_{j\neq i}^k\frac{1}{h}K^2(\frac{\bar{U}_{i\cdot}-\bar{U}_{j\cdot}}{h})(\sum_{s\neq i}^k\frac{1}{k}\eta_s)^2\frac{(k-1)^2}{k^2}\eta_j^2,\\
S_6&=\frac{2I^2}{k(k-1)}\sum_{i=1}^k\sum_{j\neq i}^k\frac{1}{h}K^2(\frac{\bar{U}_{i\cdot}-\bar{U}_{j\cdot}}{h})\frac{(k-1)^4}{k^4}\eta_i^2\eta_j^2.
\end{split}
\]
It is easy to show that $S_4=O(\frac{1}{k^2})$, $S_5=O(\frac{1}{k})$, then
\benrr
\text{E}[\hat{\Sigma}_1] = \text{E}[S_6]+O(\frac{1}{k}) = \frac{(k-1)^4}{k^4}\Sigma_1+O(\frac{1}{k}).
\eenrr
So, as $n\to \infty$, $\text{E}[\hat{\Sigma}_1]\to\Sigma_1$.

\end{subsection}

\begin{subsection}{Proof of Theorem~\ref{th:GLR1}}
Similar to the arguments in the proof of Theorem~\ref{thm:T1}, 
\benrr
\hat{a}_p(\bar{U}_{i\cdot})=a_p(\bar{U}_{i\cdot})+\eta_i,\hspace{0.3cm} i=1,...,k
\eenrr
where the new error terms $\eta_i, i=1,...,k$ are independent, $\text{E}[\eta|U]=0$, $\text{Var}[\eta|U]=\sigma_p^2(U)$.  Then we apply the GLR test for the problem without unifying the variance. 
Let $\text{RSS}_0=\sum_{i=1}^k(\hat{a}_p(\bar{U}_{i\cdot})-\hat{c})^2$ and $\text{RSS}_1=\sum_{i=1}^k(\hat{a}_p(\bar{U}_{i\cdot})-\tilde{m}_h(\bar{U}_{i\cdot}))^2$,
where  $K(\cdot)$ is a kernel function, $K_h(\cdot) = K(\cdot/h)/h$ and
\benrr
\tilde{m}_h(\bar{U}_{i\cdot})=\frac{\sum_{j=1}^kK_h(\bar{U}_{i\cdot}-\bar{U}_{j\cdot})\hat{a}_p(\bar{U}_{j\cdot})}{\sum_{j=1}^kK_h(\bar{U}_{i\cdot}-\bar{U}_{j\cdot})}.
\eenrr
Then the test statistic is given by
\benn
T_2=\frac{n}{2I}\log\frac{\text{RSS}_0}{\text{RSS}_0}\approx \frac{n}{2I}\frac{\text{RSS}_0-\text{RSS}_1}{\text{RSS}_1}.
\eenn
Then under the null hypothesis, we have
\benrr
\frac{1}{k}\text{RSS}_1 = \frac{1}{k}\sum_{i=1}^k\eta_i^2 +W_1-2W_2,
\eenrr
where
\benrr
W_1 = \frac{1}{k}\sum_{i=1}^k(\frac{\sum_{j=1}^k K_h(\bar U_{i\cdot}-\bar U_{j\cdot})\eta_j}{\sum_{j=1}^k K_h(\bar U_{i\cdot}-\bar U_{j\cdot})})^2, \quad W_2 =\frac{1}{k}\sum_{i=1}^k\eta_i\frac{\sum_{j=1}^kK_h(\bar{U}_{i\cdot}-\bar{U}_{j\cdot})\eta_j}{\sum_{j=1}^kK_h(\bar{U}_{i\cdot}-\bar{U}_{j\cdot})}.
\eenrr
It is easy to see $\sum_{i=1}^k\eta_i^2/k \rightarrow_p \int\sigma_p^2(u)f_U(u)du$ as $k \to \infty$.
To prove that $\text{RSS}_1/k$ converges to  $\int\sigma_p^2(u)f_U(u)du$ in probability,  it suffice to prove that $W_1 = o_p(1)$ and $W_2 = o_p(1)$.
Note that
\benrr
W_1 &=& \frac{1}{k^3}\sum_{i=1}^k \sum_{j=1}^k \sum_{j'=1}^k K_h(\bar U_{i\cdot} - \bar U_{j\cdot}) K_h(\bar U_{i\cdot} - \bar U_{j'\cdot}) \eta_j \eta_{j'} \frac{1}{f_U^2(\bar{U}_{i\cdot})}+o_p(1)\\
&=&\frac{1}{k^2} \sum_{j=1}^k \sum_{j'=1}^k\int K_h(u-\bar{U}_{j\cdot}) K_h(u-\bar{U}_{j'\cdot}) du \eta_j \eta_{j'}+o_p(1)\\
&=&O_p(\frac{1}{kh})+O_p(\frac{1}{kh^{1/2}}) = o_p(1).
\eenrr
Similar to the augment of $W_1$, we can see $W_2 = o_p(1)$. 
\benrr
W_2 &=& \frac{1}{k^2}\sum_{i=1}^k\sum_{j=1}^kK_h(\bar{U}_{i\cdot}-\bar{U}_{j\cdot}) \eta_i\eta_j\frac{1}{f_U(\bar{U}_{i\cdot})}+o_p(1)\\
    &=& \frac{1}{2}\frac{1}{k^2}\sum_{i=1}^k\sum_{j=1}^kK_h(\bar{U}_{i\cdot}-\bar{U}_{j\cdot}) \eta_i\eta_j(\frac{1}{f_U(\bar{U}_{i\cdot})}+\frac{1}{f_U(\bar{U}_{j\cdot})})+o_p(1)\\
    &=& O_p(\frac{1}{kh})+O_p(\frac{1}{kh^{1/2}})
\eenrr
Next we consider the term $(\text{RRS}_0-\text{RRS}_1)/k$. It can be rewritten as:
\benrr
\frac{1}{k}(\text{RRS}_0-\text{RRS}_1) &=& \frac{1}{k}\sum_{i=1}^k(\eta_i-\bar{\eta})^2-\frac{1}{k}\sum_{i=1}^k(\eta_i-(\tilde{m}_h(\bar{U}_{i\cdot})-c))^2\\
&=& W_3-W_4+O_p(\frac{1}{n})
\eenrr
where  $O_p(1/n)$ represents the term $-\bar \eta^2$  and
\benrr
W_3 = \frac{2}{k}\sum_{i=1}^k\eta_i(\tilde{m}_h(\bar{U}_{i\cdot})-c), \quad W_4 = \frac{1}{k}\sum_{i=1}^k(\tilde{m}_h(\bar{U}_{i\cdot})-c)^2.
\eenrr
First,
\benrr
W_3 &=&2\frac{1}{k}\sum_{i=1}^k\eta_i\frac{\sum_{j=1}^k K_h(\bar{U}_{i\cdot}-\bar{U}_{j\cdot}) \eta_j}{\sum_{j=1}^k K_h(\bar{U}_{i\cdot} -\bar{U}_{j\cdot})}\\
&=& \frac{2}{k^2} \sum_{i=1}^k \sum_{j=1}^k K_h(\bar U_{i\cdot} - \bar U_{j\cdot}) \eta_i \eta_j \frac{1}{f_U(\bar U_{i\cdot})}(1+o_p(1))\\
&=&(\frac{1}{k^2}\sum_{i=1}^k \frac{1}{h} K(0) \eta_i^2 \frac{2}{f_U(\bar{U}_{i\cdot})} + \frac{1}{k^2} \sum_{i=1}^k \sum_{j\neq i}^k K_h(\bar{U}_{i\cdot}-\bar{U}_{j\cdot}) \eta_i \eta_j \frac{2}{f_U(\bar{U}_{i\cdot})})(1+o_p(1))\\
&\equiv& (W_{31}+W_{32})(1+o_p(1)).
\eenrr
For $W_4$, we have
\benrr
W_4&=&\frac{1}{k}\sum_{i=1}^k(\frac{\sum_{j=1}^kK_h(\bar{U}_{i\cdot}-\bar{U}_{j\cdot})\eta_j}{\sum_{j=1}^kK_h(\bar{U}_{i\cdot}-\bar{U}_{j\cdot})})^2\\
&=&\frac{1}{k^3}\sum_{i=1}^k\sum_{j=1}^k\sum_{j'=1}^kK_h(\bar{U}_{i\cdot}-\bar{U}_{j\cdot}) K_h(\bar{U}_{i\cdot}-\bar{U}_{j'\cdot})\eta_j\eta_{j'}\frac{1}{f^2_U(\bar{U}_{i\cdot})}(1+o_p(1))\\
&=&\frac{1}{k^2}\sum_{j=1}^k\sum_{j'=1}^k\{\text{E}[K_h( U -\bar{U}_{j\cdot}) K_h( U -\bar{U}_{j'\cdot})\frac{1}{f^2_U(U)}]\} \eta_j \eta_{j'}(1+o_p(1))\\
&=&\frac{1}{k^2}\sum_{j=1}^k\sum_{j'=1}^k\frac{1}{h} \int K(v) K(v+\frac{\bar{U}_{j\cdot}-\bar{U}_{j'\cdot}}{h}) dv f_U^{-1}(\bar{U}_{j\cdot}) \eta_j \eta_{j'} (1+o_p(1)) \\
&=& (W_{41}+W_{42})(1+o_p(1))
\eenrr
where
\benrr
W_{41} &=&\frac{1}{k^2}\sum_{j=1}^k\frac{1}{h}\int K^2(v) dv f_U^{-1}(\bar{U}_{j\cdot})\eta_j^2, \\
W_{42} &=&\frac{1}{k^2}\sum_{j=1}^k\sum_{j'\neq j}^k\frac{1}{h} \int K(v) K(v+\frac{\bar{U}_{j\cdot}-\bar{U}_{j'\cdot}}{h}) dv f_U^{-1}(\bar{U}_{j\cdot})\eta_j\eta_{j'}.
\eenrr
then,
\[
\frac{1}{k}(\text{RRS}_0-\text{RRS}_1)=\Big((W_{31}-W_{41})+(W_{32}-W_{42})\Big)(1+o_p(1))+O_p(\frac{1}{n}).
\]
Note that
\[
\begin{split}
W_{31}-W_{41}&=\frac{1}{k^2}\sum_{j=1}^k \frac{1}{h}\{2K(0)-\int K^2(v)dv\}f_U^{-1}(\bar{U}_{j\cdot})\eta_j^2\\
&\to \frac{1}{kh}\{2K(0)-\int K^2(v)dv\}\int\sigma^2_p(u)du
\end{split}
\]
and
\[
W_{32}-W_{42}=\frac{1}{k^2h}\sum_{j'\neq j}\eta_j\eta_{j'}\{2K(\frac{\bar{U}_{j\cdot}-\bar{U}_{j'\cdot}}{h})-\int K(v)K(v+\frac{\bar{U}_{j\cdot}-\bar{U}_{j'\cdot}}{h})dv\}f_Y^{-1}(\bar{U}_{j\cdot}).
\]
By {\bf Proposition}~3.2 in \cite{de1987}, we have $kh^{1/2}(W_{32}-W_{42})\Rightarrow N(0, \Sigma_{W_{3,4}})$, with
\benn
\Sigma_{W_{3,4}}=2{}\int\{2K(s)-\int K(v)K(v+s)dv\}^2ds \int[\sigma^2_p(u)]^2du
\eenn
The conditions checking for {\bf Proposition}~3.2 in \cite{de1987} is almost the same with that in proof of {\bf Theorem}~5 in \cite{fan2001}, so we omit the process here.

\end{subsection}

\begin{subsection}{Proof of Theorem~\ref{thm:GLR2}}
By {\bf Theorem}~3 in \cite{zhao2018},
\benn
\frac{\widehat{\text{RSS}}_1}{k(I-p)}-\sigma^2=O_p(\frac{1}{\sqrt{n}}).
\eenn
From the definition of $T_3$, we know
\[
T_3 = \frac{n}{2}\frac{\widehat{\text{RSS}}_0-\widehat{\text{RSS}}_1}{\widehat{\text{RSS}}_1} =\frac{n}{2k(I-p)}\cdot\frac{1}{\sigma^2(1+o_p(1))}(\widehat{\text{RSS}}_0-\widehat{\text{RSS}}_1)
\]
Let $S_t$ be the index set of $\{U_{(tI-I+1)},...,U_{(tI)}\}$, and $\mbf{1}_{ti}$ is $\mbf{1}\{i\in S_t\}, t=1,...,k$.
We denote $\widehat{\bmbeta}_t=(\hat{a}_1(\bar{U}_{t\cdot}),\hat{a}_2(\bar{U}_{t\cdot}),...,\hat{a}_{p-1}(\bar{U}_{t\cdot}))^T$, $t=1, \ldots, k$,
which is the local average estimator for the first $(p-1)$ functional coefficients under $\cH_0$.
The corresponding estimator of  $\widehat{\bmbeta}_t$ under $\cH_1$ is denoted as $\widetilde{\bmbeta}$ and $\tilde{\gamma}_t=\tilde{a}_p(\bar{U}_{t\cdot})$.
To proceed further, we need more notations as:
\benrr
&&  A_t = \sum_{i\in S_t}\dot{\bX}_i \dot{\bX}_i^T, \,\,\,  B_t = \sum_{i\in S_t} X_{i,p}^2, \,\,\, C_t=\sum_{i\in S_t}\dot{\bX}_i X_{i,p}, \,\,\, D_{t1}=\sum_{i\in S_t}\dot{\bX}_i\epsilon_i \\
&& D_{t2}=\sum_{i\in S_t} X_{i,p}\ep_i, \,\,\, M_t=B_t-C_t^TA_t^{-1}C_t, \,\,\, m_{ti}=C_t^TA_t^{-1}\dot{\bX}_i-X_{i,p},
\eenrr
where  $\dot{\bX}_i = (X_{i,1}, X_{i,2}, \ldots, X_{i,p-1} ) ^T$. Note that
\benrr
\widehat{\bmbeta}_t &=& \{\sum_{i\in S_t}\dot{\bX}_i \dot{\bX}_i^T\}^{-1}  \{\sum_{i\in S_t}\dot{\bX}_i Y_i-\hat{c}\sum_{i\in S_t}\dot{\bX}_i X_{i,p}\}\\
&=&A_t^{-1} \{\sum_{i\in S_t}\dot{\bX}_i\dot{\bX}_i^T\dot{\mbf{a}}(U_i)+(c-\hat{c})\sum_{i\in S_t}\dot{\bX}_i X_{i,p}+\sum_{i\in S_t}\dot{\bX}_i\epsilon_i\}\\
&=&A_t^{-1} \{\sum_{i\in S_t}\dot{\bX}_i\dot{\bX}_i^T(\dot{\mbf{a}}(U_i) - \dot{\mbf{a}}(\bar{U}_{t\cdot})) +  A_t\dot{\mbf{a}}(\bar{U}_{t\cdot})+ C_t (c-\hat{c})+D_{t1}\},
\eenrr
where $\dot{\mbf{a}}(u)  = (a_1(u), a_2(u), \ldots, a_{p-1}(u))^T$.
Then together with {\bf Lemma}~\ref{th:localaverage}, the estimator $\widehat{\bmbeta}_t$ can be written as
\benrr
\widehat{\bmbeta}_t=\dot{\mbfa}(\bar{U}_{t\cdot})+A_t^{-1}D_{t1}+A_t^{-1}C_t(c-\hat{c})+O_p(\frac{\log n}{n}).
\eenrr
Likewise, we can have
\benrr
\widetilde{\bmbeta}_t&=&\dot{\mbf{a}}(\bar{U}_{t\cdot})+A_t^{-1}D_{t1}+A_t^{-1}C_t(c-\tilde{\gamma}_t)+O_p(\frac{\log n}{n})\\
\widehat{\bmbeta}_{t0}&=&\dot{\mbf{a}}(\bar{U}_{t\cdot})+A_t^{-1}D_{t1}+O_p(\frac{\log n}{n})
\eenrr
where $\widehat{\bmbeta}_{t0}$ is the estimator for $(a_1(\bar{U}_{t\cdot}),\ldots,a_{p-1}(\bar{U}_{t\cdot}))$ if the constant coefficient $c$ is known.
In addition, from {\bf Lemma}~\ref{th:localaverage}, we obtain
\benrr
\tilde{\gamma}_t &=& c + e_{p,p}^T\{\sum_{i\in S_t}\bX_i \bX_i^{T}\}^{-1}\{\sum_{i\in S_t} \bX_i \ep_i\}+O_p(\frac{\log n}{n})\\
&=& c+e_{p,p}^T\left[
               \begin{array}{cc}
                 A_t & C_t \\
                 C_t^T & B_t \\
               \end{array}
             \right]^{-1}\left[\begin{array}{c}
                           D_{t1} \\
                           D_{t2}
                         \end{array}\right]  +O_p(\frac{\log n}{n})\\
&=& c+(B_t-C_t^TA_t^{-1}C_t)^{-1}\left[\begin{array}{cc}
                                       -C_t^TA_t^{-1} & 1
                                     \end{array}\right]\left[\begin{array}{c}
                           D_{t1} \\
                           D_{t2}
                         \end{array}\right]+O_p(\frac{\log n}{n})\\
&=& c-M_t^{-1}\sum_{i\in S_t}m_{ti}\epsilon_i+O_p(\frac{\log n}{n}).
\eenrr
Note that $\widehat{\text{RSS}}_0-\widehat{\text{RSS}}_1$ can be expanded as
\benrr
\widehat{\text{RSS}}_0-\widehat{\text{RSS}}_1 &=& \sum_{t=1}^k \sum_{i=1}^n (Y_i-\widehat{\bmbeta}_t^T \dot{\bX}_i -\hat{c} X_{i,p})^2\mbf{1}_{ti} - \sum_{t=1}^k \sum_{i=1}^n (Y_i-\widetilde{\bmbeta}_t^T \dot{\bX}_i-\tilde{\gamma}_t X_{i,p})^2\mbf{1}_{ti}\\
&=&\sum_{t=1}^k \sum_{i=1}^n \{(Y_i-\widehat{\bmbeta}_t^T \dot{\bX}_i -\hat{c}X_{i,p})^2- (Y_i-\widehat{\bmbeta}_{t0}^T \dot{\bX}_i -c X_{i,p})^2\}\mbf{1}_{ti}\\
&&+\sum_{t=1}^k\sum_{i=1}^n\{(Y_i-\widehat{\bmbeta}_{t0}^T \dot{\bX}_i -c X_{i,p})^2-(Y_i-\widetilde{\bmbeta}_t^T \dot{\bX}_i - \tilde{\gamma}_t X_{i,p})^2\}\mbf{1}_{ti}\\
&\equiv&\text{DRSS}_1+\text{DRSS}_2.
\eenrr
We first consider $\text{DRSS}_1$, which can be rewritten as:
\benrr
&&\sum_{t=1}^k \sum_{i=1}^n \{\dot{\bX}_i^T(\widehat{\bmbeta}_{t0}-\widehat{\bmbeta}_{t})+X_{i,p}(c-\hat{c}) \} \{2Y_i-\widehat{\bmbeta}_t^T \dot{\bX}_i -\widehat{\bmbeta}_{t0}^T \dot{\bX}_i-\hat{c} X_{i,p}- c X_{i,p}\}\mbf{1}_{ti}\\
&=&\sum_{t=1}^k\sum_{i=1}^n\{\dot{\bX}_i^T(\widehat{\bmbeta}_{t0}-\widehat{\bmbeta}_{t}) + X_{i,p}(c-\hat{c}) \} \{2\mbf{\dot{a}}^T(U_i)\dot{\bX}_i+c X_{i,p}-\widehat{\bmbeta}_t^T \dot{\bX}_i -\widehat{\bmbeta}_{t0}^T \dot{\bX}_i-\hat{c} X_{i,p}\}\mbf{1}_{ti}\\
&=&\sum_{t=1}^k\sum_{i=1}^n\{\dot{\bX}_i^T(\widehat{\bmbeta}_{t0}-\widehat{\bmbeta}_{t})+X_{i,p}(c-\hat{c}) \} \{\dot{\bX}_i^T(2\mbf{\dot{a}}(U_i)-\widehat{\bmbeta}_t-\widehat{\bmbeta}_{t0}) +X_{i,p}(c-\hat{c})\}\mbf{1}_{ti}.
\eenrr
Therefore $\text{DRSS}_1= \text{DRSS}_1^*(1+o_p(1))$ where
\benrr
\text{DRSS}_1^* = \sum_{t=1}^k\sum_{i=1}^n m_{ti}(\hat{c}-c)\times\{m_{ti}(\hat{c}-c)-2\dot{\bX}_i^TA_t^{-1}D_{t1}+2\epsilon_i\}\mbf{1}_{ti}.
\eenrr
Note that $m_{ti} = (C_t^T A_t^{-1} \dot{\bX}_i - X_{i,p})$. Then $\text{DRSS}_1^*$ can be further written as
\benrr
\text{DRSS}_1^*&=& (\hat{c}-c)^2\sum_{t=1}^kM_t+2(\hat{c}-c)\sum_{t=1}^k\sum_{i\in S_t}m_{ti}\epsilon_i.
\eenrr
By the law of large numbers, $(1/n)\sum_{t=1}^k\sum_{i\in S_t}m_{ti}\epsilon_i=O_p(1/\sqrt{n})$.
In addition, $\hat{c}-c=O_p(1/\sqrt{n})$, $\sum_{t=1}^kM_t=O_p(n)$. Hence $\text{DRSS}_1=O_p(1)$.

Next we deal with  $\text{DRSS}_2$. Note that
\benrr
\text{DRSS}_2&=&\sum_{t=1}^k\sum_{i=1}^n\{\dot{\bX}_i^T(\widetilde{\bmbeta}_{t}-\widehat{\bmbeta}_{t0})+X_{i,p}(\tilde{\gamma}_t-c) \} \\
&&\times\{\dot{\bX}_i^T(\mbf{\dot{a}}(U_i)-\widetilde{\bmbeta}_t)+\dot{\bX}_i^T(\mbf{\dot{a}}(U_i)-\widehat{\bmbeta}_{t0}) +X_{i,p}(c-\tilde{\gamma}_t)\}\mbf{1}_{ti}.
\eenrr
Then, by Lemma~1,  $\text{DRSS}_2 = \text{DRSS}_2^*(1+o_p(1))$ where
\benrr
\text{DRSS}_2^*=\sum_{t=1}^k\sum_{i=1}^n -m_{ti}(\tilde{\gamma}_t-c)\times\{m_{ti}(\tilde{\gamma}_t-c)-2\dot{\bX}_i^TA_t^{-1}D_{t1}+2\epsilon_i\}\mbf{1}_{ti}
\eenrr
Further
\benrr
\text{DRSS}_2^* &=& -\sum_{t=1}^kM_t(\tilde{\gamma}_t-c)^2-2\sum_{t=1}^k\sum_{i\in S_t}(\tilde{\gamma}_t-c)m_{ti}\epsilon_i\\
&=&\sum_{t=1}^kM_t^{-1}\sum_{i\in S_t}\sum_{j\in S_t}m_{ti}m_{tj}\epsilon_i\epsilon_j = P_1 + P_2
\eenrr
where
\benn
P_1 = \sum_{t=1}^kM_t^{-1}\sum_{i\in S_t}m_{ti}^2\epsilon_i^2, \quad P_2= \sum_{t=1}^kM_t^{-1}\sum_{i\in S_t}
\sum_{\substack{{j\in S_t}\\{j \neq i}}}m_{ti}m_{tj}\epsilon_i\epsilon_j.
\eenn
Now we shall show that
\benn
\frac{P_1-k\sigma^2}{\sqrt{v_1}} \Rightarrow N(0,1), \quad \frac{P_2}{\sqrt{v_2}} \Rightarrow N(0,1),
\eenn
where
\benn
v_1=(\mu_4-\sigma^4)\sum_{t=1}^k M_t^{-2}\sum_{i\in S_t}m_{ti}^4, \quad v_2=2k\sigma^4-2\sigma^4 \sum_{t=1}^k M_t^{-2} \sum_{i\in S_t}m_{ti}^4.
\eenn
Since
\benn
P_1-k\sigma^2=\sum_{t=1}^kM_t^{-1}\sum_{i\in S_t}m_{ti}^2\epsilon_i^2-k\sigma^2=\sum_{t=1}^kM_t^{-1}\sum_{i\in S_t}m_{ti}^2(\epsilon_i^2-\sigma^2),
\eenn
let $Z_t=M_t^{-1}\sum_{i\in S_t}m_{ti}^2(\epsilon_i^2-\sigma^2)$, then $Z_t$ is independent random variable and
\benn
\text{E}[Z_t]=0, \hspace{1cm}\text{Var}[Z_t]=(\mu_4-\sigma^4)M_t^{-2}\sum_{i\in S_t}m_{ti}^4
\eenn
Because $\frac{1}{2}\leq M_t^{-2}\sum_{i\in S_t}m_{ti}^4\leq1$ for any $t$, and $v_1=\sum_{t=1}^k\text{Var}[Z_t]=O(k)$, it is easy to prove that the Lindeberg's condition is satisfied for $Z_t$.
Therefore, by central limit theorem,
\benn
\frac{P_1-k\sigma^2}{\sqrt{v_1}} = \frac{\sum_{t=1}^kZ_t}{\sqrt{v_1}} \Rightarrow N(0,1).
\eenn
The proof of the asymptotic normality of $P_2$ is an application of {\bf Proposition} 3.2 in \cite{de1987}. 
Denote
\benn
\Pi_{ij}=\sum_{t=1}^k\mbf{1}_{ti}\mbf{1}_{tj}, i.e.,  \Pi_{ij}=
\begin{cases}
 1 & i \ \text{and}\  j\  \text{are in the same group}\\
 0 & \text{otherwise}\\
 \end{cases}
\eenn
Define $W_{ij}=2M_t^{-1}m_{ti}m_{tj}\Pi_{ij}\epsilon_i\epsilon_j$, then $P_2=\sum_{i<j}W_{ij}$. Then by the {\bf Proposition} 3.2 in \cite{de1987}, it suffice to check the following conditions:
\begin{enumerate}
  \item $\text{E}[W_{ij}|\epsilon_i]=0$ a.s.  for all $i,j \leq n$.
  \item $\text{Var}[P_2]\to v_2$.
  \item $G_{\RN{1}}, G_{\RN{2}}, G_{\RN{4}}$ is of smaller order than $v_2^2$.
\end{enumerate}
where
\benrr
G_{\RN{1}}&=&\sum_{i<j}\text{E}[W_{ij}^4] , \quad G_{\RN{2}}=\sum_{i<j<m}\text{E}[W_{ij}^2W_{im}^2+W_{ji}^2W_{jm}^2+W_{mi}^2W_{mj}^2],\\
G_{\RN{4}}&=&\sum_{i<j<m<l}\text{E}[W_{ij}W_{im}W_{lj}W_{lm}+W_{ij}W_{il}W_{mj}W_{ml}+W_{im}W_{il}W_{jm}W_{jl}].
\eenrr
Condition 1 is obvious by the definition. To prove condition 2, note that $\text{E}[P_2]=0$, then
\benrr
&&\text{Var}[P_2]=\text{E}[P_2^2] =\text{E}[(\sum_{t=1}^kM_t^{-1}\sum_{i\in S_t}\sum_{\substack{{j\in S_t}\\{j \neq i}}}m_{ti}m_{tj}\epsilon_i\epsilon_j)^2]\\
&=&\sum_{t=1}^k M_t^{-1} \text{E}[(\sum_{i\in S_t}\sum_{\substack{{j\in S_t}\\{j \neq i}}}m_{ti}m_{tj}\epsilon_i\epsilon_j)^2] = 2\sigma^4\sum_{t=1}^kM_t^{-1}\sum_{i\in S_t}m_{ti}^2\sum_{\substack{{j\in S_t}\\{j \neq i}}}m_{tj}^2\\
&=&2\sigma^4\sum_{t=1}^kM_t^{-2}(\sum_{i\in S_t}m_{ti}^2)^2-2\sigma^4\sum_{t=1}^kM_t^{-2}\sum_{i\in S_t}m_{ti}^4 = 2k\sigma^4-2\sigma^4\sum_{t=1}^kM_t^{-2}\sum_{i\in S_t}m_{ti}^4
\eenrr
So Condition 2 is satisfied and we obtain $v_2^2=O(k^2)$.
For Condition 3,
\benrr
G_{\RN{1}}&=&\sum_{i<j}\text{E}[W_{ij}^4]=\sum_{i<j}\text{E}[(2M_t^{-1}m_{ti}m_{tj}\Pi_{ij}\epsilon_i\epsilon_j)^4]\\
&= &8\mu_4^2\sum_{t=1}^kM_t^{-4}\sum_{i\in S_t}m_{ti}^4\sum_{\substack{{j\in S_t}\\{j \neq i}}}m_{tj}^4 =O(k)
\eenrr
Similarly, we can prove that $G_{\RN{2}}=O(k)$, $G_{\RN{4}}=O(k)$.

Combining the asymptotic results of $P_1$ and $P_2$, we have
\[
\frac{\text{DRSS}_2-k\sigma^2}{\sqrt{v_3}}\Rightarrow N(0, 1)
\]
where $v_3=v_1+v_2+2\text{Cov}(P_1,P_2)$. It is easy to prove that $\text{Cov}(P_1,P_2)=0$. Then
\[
v_3=(\mu_4-3\sigma^4)\sum_{t=1}^k\text{E}[M_t^{-2}\sum_{i\in S_t}m_{ti}^4]+2k\sigma^4=(\mu_4-3\sigma^4)\Psi_n+2k\sigma^4
\]
Since $v_3$ is of order $O(k)$, then we have
\[
v_3^{-1/2}(\widehat{\text{RSS}}_0-\widehat{\text{RSS}}_1-k\sigma^2)\to N(0,1)
\]
Let $\sigma_3^2=(\frac{\mu_4}{\sigma^4}-3)\Psi_n+2k$, then
\[
\frac{2(I-p)}{I}\sigma_3^{-1}(T_3-\frac{n}{2(I-p)})\to N(0,1)
\]

\end{subsection}







\vskip 0.2in
\bibliography{LAPE}

\end{document}